\documentclass[10pt]{article}

\usepackage{amsmath}
\usepackage{amssymb}

\usepackage{graphicx}

\usepackage{cite}

\usepackage{color} 

\usepackage[labelfont=bf,labelsep=period,justification=raggedright]{caption}

\usepackage[parfill]{parskip}

\makeatletter
\renewcommand{\@biblabel}[1]{\quad#1.}
\makeatother

\date{}

\renewcommand{\sp}{\text{sp}}

\begin{document}

% Title must be 150 characters or less
\begin{flushleft}
{\Large
\textbf{Adaptive probabilistic neural coding from deterministic spiking neurons: analysis from first principles}
}
% Insert Author names, affiliations and corresponding author email.
\\
Michael Famulare$^{1,\ast}$, 
Adrienne Fairhall$^{2}$
\\
\bf{1} Department of Physics, University of Washington, Seattle WA, USA
\\
\bf{2} Department of Physiology and Biophysics, University of Washington, Seattle WA, USA
\\
$\ast$ E-mail: famulare@uw.edu
\end{flushleft}

% Please keep the abstract between 250 and 300 words 
\section*{Abstract}
A neuron transforms its input into output spikes, and this transformation is the basic unit of computation in the nervous system.  The spiking response of the neuron to a complex, time-varying input can be predicted from the detailed biophysical properties of the neuron, modeled as a deterministic nonlinear dynamical system. In the tradition of neural coding, however, a neuron or neural system is treated as a black box and statistical techniques are used to identify functional models of its encoding properties. The goal of this work is to connect the mechanistic, biophysical approach to neuronal function to a description in terms of a coding model.  Building from preceding work at the single neuron level, we develop from first principles a mathematical theory mapping the relationships between two simple but powerful classes of models: deterministic integrate-and-fire dynamical models and linear-nonlinear coding models.  To do so, we develop an approach for studying a nonlinear dynamical system by conditioning on an observed linear estimator. We derive asymptotic closed-form expressions for the linear filter and estimates for the nonlinear decision function of the linear/nonlinear model. We analytically derive the dependence of the linear filter on the input statistics and we show how deterministic nonlinear dynamics modulate the properties of a probabilistic code.  We demonstrate that integrate-and-fire models without any additional currents can perform perfect contrast gain control, a sophisticated adaptive computation, and we identify the general dynamical principles responsible. We then design from first principles a nonlinear dynamical model that implements gain control.   While we focus on the integrate-and-fire models for tractability, the framework we propose to relate LN and dynamical models generalizes naturally to more complex biophysical models.

% Please keep the Author Summary between 150 and 200 words
% Use first person. PLoS ONE authors please skip this step. 
% Author Summary not valid for PLoS ONE submissions.   
\section*{Author Summary}
One of the primary goals of sensory neurophysiology is to characterize the stimulus information that is extracted and transmitted by a neural system. This is typically done by identifying stimulus features to which the system is sensitive, and then finding the probability of response as a function of those stimulus features.  Together, the relevant features and response define a coding model.  Finding a correspondence between coding models and underlying neuronal circuitry and biophysics has been elusive. Here, we derive the relationship between such a coding model and biophysical parameters in the case of an important class of single neuron models.  This derivation allows us explore a remarkable property of neural coding: the ability of neural systems to represent information in relative units. We find parameters for which this property may be perfectly realized by an individual neuron.

\section*{Introduction}
The neuron is arguably the fundamental information processing element of the nervous system, encoding a pattern of inputs into a sequence of output spikes.  To what properties of the input is the neuron sensitive?  What mathematical form does the transformation from input to spiking output take?  What are the general principles governing that transformation?   These coding questions are paired with questions of implementation: what biophysical mechanisms support the observed computational properties?  Of the hundreds of known ion channels expressed throughout nervous systems, why do channels with specific kinetics appear with particular densities in particular locations?  How do the biophysical parameters of neurons map onto their computational function? To start to address these questions, our goal in this work is to develop an explicit analytic bridge between a dynamical systems description of neuronal behavior and a coding model.

Integrating a biophysical and computational perspective on single neuron function requires an integration of the distinct mathematics of dynamics and statistics.  On the coding side, the computational framework that we study in this paper is that of the linear-nonlinear (LN) model. The LN model is defined by a linear filter representing the feature that is relevant to firing a spike, and a decision function that acts on the filtered stimulus to determine the instantaneous firing rate. For a white noise input with fixed mean and variance, the unbiased and consistent LN model given the input statistics can be found with reverse correlation techniques. When the relevant feature space is low-dimensional, the LN model constitutes a complete description of the neuron's computation when the instantaneous firing rate is the only relevant output statistic\cite{Aguera2003a}. 

In contrast, the natural language of neuronal biophysics is nonlinear dynamics.  In the limit of large numbers of channels, the neuron's behavior is described by a potentially high-dimensional set of nonlinear differential equations for the evolution of the membrane potential and the state of the ion channels as a function of current input\cite{GerstnerKistler}.  Spikes are stereotyped events in the time series of the membrane voltage.  We will limit ourselves here to single-compartment models, and to the case in which internal noise is negligible and so spikes are generated deterministically in response to current input. 

Although one would like to understand the general relationship between a multivariate dynamical model and an LN coding model, for tractability we will start with models with a single dimension that nonetheless incorporate some of the properties of conductance-based models, the leaky integrate-and-fire (LIF) and the exponential integrate-and-fire (EIF) models.  The LIF model describes the dynamics of the voltage as linear below a voltage threshold, at which spikes are generated instantaneously. This model is effective when there is a large separation of timescales between those of the sodium and potassium channel dynamics responsible for spike generation and the passive ``leak" voltage dynamics of the membrane (which includes the membrane properties and any linear contributions to the subthreshold dynamics). The EIF model adds to this an exponential subthreshold nonlinearity, providing an intrinsic instability that leads to spiking. This model has been shown to fit well to data from cortical neurons\cite{Badel2008b,Badel2008c} and the Wang-Buz\'{a}ki model of hippocampal fast spiking interneurons\cite{Wang1996,Fourcaud2003,Badel2008c}.  

A key property of neural systems is that they adapt to the statistics of the input\cite{Barlow1961,Laughlin1981,Laughlin1989,Smirnakis1997}. Single neuron computation as characterized by LN models also generally appears \emph{adaptive} in the sense that the optimal LN model for predicting neuronal response changes as the statistics of the input change\cite{Brenner2000b,Fairhall2001,Yu2003,Yu2005a,Wark2007}.  We will address this property explicitly, focusing on changes in stimulus variance.  As the variance of the white noise input increases, the optimal filters tend to have shorter timescales and the decision function changes slope, or gain\cite{Shapley1978,Kim2001,Baccus2002,Maravall2007}. We consider in this paper the phenomenon of perfect \emph{contrast gain control}, whereby the decision function adapts to changes in the input standard deviation (contrast) such that the fine temporal structure of the firing rate is controlled by the size of the input relative to the input standard deviation.   Biophysical models, on the other hand, do not adapt in the sense given above: the model does not change form when the input statistics are changed.   While slow dynamics may cause spike frequency changes, these will be included in a complete biophysical model of a neuron via differential equations with fixed parameters. We aim to determine under what conditions a {\em fixed} neuronal dynamical model leads to perfect contrast gain control.

In this paper, we accomplish two primary goals.  First, we demonstrate how to derive LN models from deterministic dynamical models from first principles.  The two models are related by a voltage estimation problem: the linear filter estimates the state of the membrane voltage, and the nonlinear decision function is determined by the interaction of the intrinsic spike-generating currents and the precision of the linear voltage estimate.  We propose that the proper choice of filter maximizes the precision of the estimate of the voltage in the spike-generating regime. We identify the principles behind why the optimal filter adapts to changes in input standard deviation, and we derive approximate expressions for the form of the optimal filter in various limits.  From the perspective of this first goal, the deterministic spiking model is taken to be fundamental.  The optimal LN model provides a representation of the \emph{encoding} performed by the spiking model but, as an approximation, is necessarily stochastic.

Second, we change perspectives and take the family of LN models to be fundamental and assume that they completely describe the relevant computation of the neuron.  From this perspective, the dynamical details that the LN model does not predict can provide a mechanism to modulate the optimal \emph{decoding} of a single spike in response to changes in contextual properties of the stimulus. 
While for small input standard deviations, the LIF model is a detector of threshold crossings, we show that for large standard deviations, the LIF model generically shows perfect contrast gain control.  We identify the general principles behind how simple dynamical models can perform this ``intelligent" adaptive computation. We then show that this general understanding of contrast gain control can be used to constrain kinetic parameters in the EIF model to allow it to show contrast gain control without requiring asymptotically large input standard deviation.  

The framework we propose to relate LN and dynamical models is quite general. We focus on the integrate-and-fire models because they are simple, surprisingly rich, and allow us to obtain analytic results that make more general points. However, the fact that more complex neurons can often be reduced to an EIF model means that our results have power beyond this simple case. We will discuss how this framework extends to higher-dimensional neuronal models.

% Results and Discussion can be combined.
\section*{Results}
We drove integrate-and-fire models with white noise current with zero mean and standard deviation proportional to $\sigma$, where $\sigma$ is reported in units of the difference between the resting and threshold voltages, $v_{th}-v_o$ (see Section: {\em Integrate-and-fire models}).   Since the mean current can be absorbed into the resting and threshold potentials, we fix it to zero and focus only on changes in the input standard deviation (``contrast'').  Note that while the inputs are drawn randomly, they are completely specified---we do not consider any unknown background noise.  Thus the response of the dynamical model to the input is deterministic and the instantaneous firing rate, $R(t)$, is either zero or $dt^{-1}$, where $dt$ is the sampling time step.  

For each input standard deviation, we characterized the computation performed by the deterministic dynamical model with an LN model.  As explained in Section: {\em Identifying LN models}, an LN model for a single neuron produces an estimate of the instantaneous firing rate in response to the input current.  The LN model consists of two parts: a feature, $h(t)$, that acts on the input to produce a linearly filtered stimulus, $s(t)$, which is the amplitude of the feature present in the input, and a nonlinear threshold function that acts on the filtered stimulus to estimate the instantaneous firing rate.  

For Gaussian white noise inputs, the spike-triggered average current (STA), defined in Eq. \eqref{STA}, is the consistent and unbiased choice for the filter; the STA of the LN model is the STA of the dynamical model. However, the STA is not necessarily the \emph{optimally predictive} filter.  We examined three choices of the filter: the normalized STA, which we denote $h_s$; the membrane filter, $h_m$, introduced in Eq. \eqref{MembraneFilter}; and the stochastic linearization filter, $h_l$, introduced in Eq. \eqref{VoltageFilter}. The filtered stimulus $s_x(t)$ is defined as the convolution of the input current with the filter with corresponding subscript, $h_x$. The threshold function that predicts the firing rate---the rate estimation function---is defined relative to a specific choice of filter. The filter identification is denoted simply through its dependence on $s_x$: $R_{\sigma}[s_x]$.

\subsection*{LN models for the LIF model based on the spike-triggered average filter}
\begin{figure}[!ht]
\begin{center}
\includegraphics{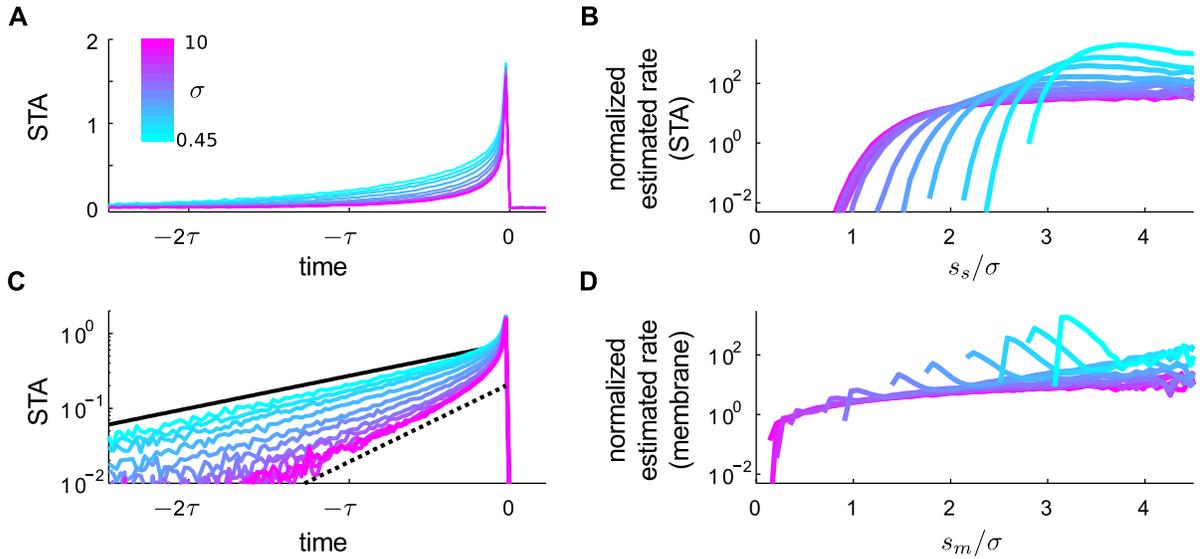}
\end{center}
\caption{ {\bf LN models for the LIF model} for $0.45 \le \sigma \le 10$.  \textbf{A} . Spike-triggered average in units of $\frac{\langle i(t)|\sp\rangle}{\sigma\sqrt{\tau/{dt}}}$. The color code represents input standard deviation and is preserved across all figures unless otherwise indicated. \textbf{C}. STA on log scale.  The solid black line is the passive membrane filter in Eq. \eqref{MembraneFilter}. Dotted is the stochastic linearization filter in the large-$\sigma$ limit (see Eq. \eqref{KslLIF}). The STA adapts to changes in the input statistics and approaches a fixed form for large $\sigma$. \textbf{B}. Simulation results for the normalized rate estimation function based on the STA filter, $R_{\sigma}\!\left[s_s\right]/\bar{R}_{\sigma}$ corresponding to Eq. \eqref{NonLinFromData}; data is scaled by the mean firing rate and plotted with respect to relative stimulus strength, $s_s/\sigma$.    Rate estimation functions are graded thresholds.  For large $\sigma$, the LN models show perfect contrast gain control as defined in Eq. \eqref{NonLinCGC}. \textbf{D}. As in B, but for the LN models based on the passive membrane filter.  Comparison of C and D supports the results about the predictive power of different filters at different limits in Fig. \protect \ref{CoinFactFig}, as discussed in the text.  At low $\sigma$, the rate estimation functions for the membrane filter are more selective (higher stimulus threshold) and are more precise (sharply-peaked) than the STA-based models, and are thus more informative, as argued preceding Eq. \eqref{HOptLowSigLIF}.  For large $\sigma$, the STA-based model is more selective and has a larger dynamic range, giving it greater predictive power in comparison to the membrane filter models, as anticipated by the argument preceding Eq. \eqref{HOptHighSigLIF}.
} 
\label{LNLIFFig}
\end{figure}

Spike times in the LIF model correspond to the instants when the membrane voltage is above threshold: $v(t) \ge v_{th}$.  We used reverse correlation techniques (see Section: {\em Identifying LN models with reverse correlation}) to build LN models in which the filter is the spike-triggered average current (STA) defined in Eq. \eqref{STA}.    Simulations were run with the resting and reset voltages equal: $v_{r}=v_o$. The results for the LIF model are qualitatively insensitive to the values of the parameters used.  Results are shown in Fig. \ref{LNLIFFig}A--C for $0.45 \le \sigma \le 10$. 

The STA of an LIF neuron when driven by stimuli with different variances has been fairly well-studied, numerically and analytically \cite{Aguera2003b,Paninski2006IFSTA}. The STA is an integrating filter for all input standard deviations.  The detailed shape of the STA can be decomposed into an exponential tail at longer times and a rapid rise at short times, as shown in Eq. \eqref{StaAnDt} and Fig. \ref{LIFStaAnFig}.  Both components depend on the input standard deviation.  In all cases, at times very near the spike time and short compared to the membrane time constant, $\tau$, the STA shows a rapid upward fluctuation whose peak value is proportional to $\sigma$ and depends on the correlation time of the input (see Section: {\em Discretization and regularization} for discussion).    This singular component of the STA is associated with crossing threshold from below \cite{Aguera2003b,Hong2007,Burak2009} and its detailed shape is due to the discontinuous dynamics of the spike \cite{Paninski2006IFSTA,Badel2006,Badel2008a}.  The longer-time behavior of the STA is exponential.  For small $\sigma$ only, the timescale of the exponential is given by the membrane time constant, $\tau$.  For finite $\sigma$, the timescale of the STA is always less than the membrane time constant. At large $\sigma$, the STA goes to a fixed form---the normalized STA becomes invariant to changes in the input standard deviation. 

Rate estimation functions were built using Eq. \eqref{NonLinFromData} based on the STA-filtered stimulus, $s_s(t)$.  Qualitatively, the rate estimation function based on the STA-filtered stimulus is a graded threshold, with a form that depends on $\sigma$.  As $\sigma$ is increased, there is a shift in the minimum value of the filtered stimulus that permits spiking (a so-called ``subtractive'' gain change). For small $\sigma$, the rate estimation function is non-monotonic, and first becomes non-zero near a threshold, $s_{th}=\sqrt{2}(v_{th}-v_o)$.  For large $\sigma$, the rate estimation function increases monotonically and becomes linear in $s_s$ for large $s_s$. In terms of the relative stimulus strength, ${s_s}/{\sigma}$, the rate estimation function shows \emph{perfect} contrast gain control as defined in Eq. \eqref{NonLinCGC}: after scaling out a multiplicative gain factor---the mean firing rate---the threshold function is invariant to changes in the input standard deviation.

Thus, the STA-based LN models describe the encoding of the LIF model as leaky integration followed by thresholding; furthermore the encoding is \emph{adaptive}, with both the STA and the threshold function depending on the input standard deviation. We will return to the observation of perfect contrast adaptation in Section: {\em Context-dependent coding in integrate-and-fire models}.

\subsection*{Derivation of rate estimation functions from the voltage}
Our first goal is to derive the components of the LN model, as sampled numerically from a dynamical model, directly from that underlying model. In the special case of single neurons, we can use the fact that spikes are defined in terms of voltage state.  Previous work has focused on how the filter is determined by the dynamics\cite{Kistler1997,Aguera2003a,Paninski2006IFSTA,Paninski2006STV,Hong2007,Burak2009,Famulare2010}. Now, we will first derive expressions for the threshold function of the LN model for a generic filter, which we call here the rate estimation function, and then return to the question of the appropriate filter.
 
The fundamental insight of this work is to recognize that the estimated firing rate produced by an LN model can be derived from the dynamics of the voltage, by averaging over all voltage trajectories whose evolution is consistent with the occurrence of a spike at time $t$ and the filtered stimulus taking on the observed value, $s_x(t)$ for filter $h_x$:
\begin{equation}
R_{\sigma}\!\left[s_x(t)\right]=\displaystyle \int\! \mathcal{D}v(t')\,  p_{\sigma}\!\left[v(t')\big|s_x(t)\right] \,R[v(t')]. \label{PathEq} 
\end{equation}
The integral is over all voltage trajectories, weighted by the probability that the trajectory, $v(t')$, is consistent with the value of the filtered stimulus at time $t$, $p_{\sigma}\!\left[v(t')\big|s_x(t) \right]$. The term $R[v(t')]$ represents the instantaneous firing rate at time $t$ for a given voltage trajectory; it is non-zero only at events in the voltage trajectory that are defined to be spike times.   Thus, mapping the dynamics onto the coding model requires (1) specifying the mapping from a voltage trajectory to unique spike times, and (2) understanding the relationship between the voltage and the filtered stimulus.  

\subsubsection*{Rate estimation functions for the LIF model}
For the leaky integrate-and-fire neuron, the transformation from voltage to spikes is simply via threshold-crossing, and the rate is given in terms of the voltage by
\begin{equation}
R[v(t)]=\frac{1}{dt}\text{H}\!\left[v(t)-v_{th}\right], \label{RofVLIF}
\end{equation}
where $dt$ is the sampling time step, $v_{th}$ is the threshold voltage, and $\text{H}[\ldots]$ is the Heaviside step function; i.e. for time bins $dt$, spikes occur with probability one when the voltage is above $v_{th}$ and zero otherwise; see Section: {\it Discretization and regularization} for relevant discussion.  
Then Eq. \eqref{PathEq} simplifies to 
\begin{align}
R_{\sigma}\!\left[s_x(t)\right] = \displaystyle \int \! dv(t)\,R[v(t)]\,p_{\sigma}\!\left[v(t)\big|s_x(t)\right]. \label{NonLinFromVoltageLIF0}
\end{align}
Eqs. \eqref{RofVLIF} and \eqref{NonLinFromVoltageLIF0} combine to give:
\begin{align}
R_{\sigma}\!\left[s_x(t) \right] &= \frac{1}{dt}\displaystyle \int_{v_{th}}^{\infty} dv(t) \,p_{\sigma}\!\left[v(t) \big|s_x(t) \right], \notag\\
&=\frac{1}{dt}P_{\sigma}\!\left[v(t)\ge v_{th}\big|s_x(t)\right], \label{NonLinFromVoltageLIF}
\end{align}
where $P_{\sigma}\!\left[v(t) \ge v_{th}\big|s(t) \right]$ is the probability that the voltage is above threshold given the value of the filtered stimulus (throughout this work, we use the convention that probability densities are denoted with lowercase $p$ and true probabilities are denoted with uppercase $P$).  

To proceed, we need to analyze $p_{\sigma}\!\left[v(t)\big|s_x(t) \right]$. The form of Eq. \eqref{NonLinFromVoltageLIF} suggests that the filtered stimulus acts as a linear estimator of the state of the voltage at the time of a spike.  While no general closed form solution is available for this voltage estimation problem, in Section: {\em Moment-based asymptotic results}, we derive useful asymptotic results at small and large $\sigma$.  First, however, we discuss the exponential integrate-and-fire model and the question of optimal filter choice.

\subsubsection*{Rate estimation functions of the EIF model}

While spike times are very clearly specified in terms of the voltage for the leaky integrate-and-fire model, this is not generally the case for more realistic neurons; the spike is an extended waveform that is stereotyped but always shows some variability.  This is true even for the exponential integrate-and-fire model: only the peak of the spike is uniquely specified and there is variability in the spike onset.  To identify an equivalent function relating rate to voltage in this more realistic case, we must identify an effective threshold above which the spike waveform is stereotyped.

For inputs whose standard deviation goes to zero, the EIF model has an intrinsic voltage threshold whose crossing times provide the natural definition for the onset time of the spike.  The unstable fixed point, $v_{th}$, is the separatrix between subthreshold and spiking trajectories, known as the \emph{dynamical threshold} \cite{Azouz2000,Izhikevich2001,GerstnerKistler,Hong2007,Famulare2010}. At low input noise, the stimulus brings the system to and across the dynamical threshold, above which the spiking trajectory is essentially independent of the stimulus. The spike-generating dynamics are controlled by two parameters: the voltage activation scale for the exponential current, $\Delta$, and the after-spike reset voltage, $v_r$ (see Eq. \ref{LIFModel}).

For larger input standard deviations, there is a significant probability that trajectories that cross the dynamical threshold may cross back without spiking, and so we need a more appropriate definition of threshold.   To identify LN models that best represent how the input drives spiking, we want to identify spike times that separate the role of subthreshold integration from super-threshold, largely input-independent dynamics. We define a \emph{stochastic dynamical threshold}, $v_{th,\sigma}$ to be the voltage beyond which fluctuations in the input that occur immediately following the crossing point have a fixed, low probability of aborting the voltage from reaching the spike peak:
\begin{equation}
C=\displaystyle \int_{v_{th,\sigma}}^{\infty} dv(t+dt)\, p_{\sigma}\!\left[v(t+dt)\big|v_{th,\sigma}\right]. \label{StochasticThreshold}
\end{equation}
Here C is a number between 0.5 and 1, and the transition probability from time $t$ to $t+dt$ is given in Eq. \eqref{FreeProcess}. While this rule only considers the instantaneous evolution of the voltage at the threshold, the model's exponential nonlinearity ensures that future time steps above threshold are exponentially more likely to continue to the top of a spike. So, for an appropriate choice of $C$ (to be discussed shortly), this local definition identifies spikes reliably.  Integrating Eq. \eqref{StochasticThreshold} gives an implicit equation for the threshold as a function of the noise strength:
\begin{equation}
v_o-v_{th,\sigma}+f\left(v_{th,\sigma}\right)=\sigma\sqrt{\frac{2\tau}{dt}}\text{erf}^{-1}\!\left(2C-1\right). \label{StochasticThreshold2}
\end{equation}
There are two roots to this equation; the threshold is the one such that $v_{th,\sigma}\ge v_{th}$.  The unstable fixed point $v_{th}$ is recovered for C = 0.5: it is the voltage from which all perturbations have an equal chance of either stepping toward the spike peak or toward the stable fixed point in the next time step.  For corrections to the threshold that are small relative to the activation scale $\Delta$, the stochastic dynamical threshold is
\begin{equation}
v_{th,\sigma}\approx v_{th}+\sigma\sqrt{\frac{2\tau}{dt}}\frac{\text{erf}^{-1}\!\left(2C-1\right)}{f'\!\left(v_{th}\right)-1}, \label{StochasticThreshold3}
\end{equation}
which is linear in the standard deviation.  The position of the stochastic dynamical threshold is controlled by the spike-generating dynamics, and is determined by the magnitude of the deterministic excitable current needed to compensate for input fluctuations that work to abort a spike.

As discussed in Section: {\em Discretization and regularization}, the $\sqrt{dt^{-1}}$ divergence of the stochastic dynamical threshold arises from the white noise statistics of the input.  If the continuous time limit is taken without regularization, the derivative of the voltage can be arbitrarily large and so no finite voltage threshold satisfies the criterion in Eq. \eqref{StochasticThreshold}. This expression is sensible for inputs correlated on the timescale $dt$, provided $dt$ is small. 

To construct an LN model, we need to specify the confidence, $C$, that determines the threshold used to identify spikes. In a real neuron, a natural definition of threshold corresponds to the peak voltage of the minimal event that propagates down the axon, and $C$ should be set accordingly. Here, the choice of $C$ is somewhat arbitrary.  The threshold should identify event times that correspond to true spikes (those that reach $v_s$) with a low fraction of mislabeled aborted spikes.  This criterion implies $C\rightarrow 1$, corresponding to triggering spikes at the peak voltage.  However, for the purposes of constructing a model that best characterizes the relationship of spiking to the input, one would like to set this threshold as low as possible, in order to minimize the relative effect of the stimulus-independent internal spike-generating dynamics.  This would imply $C\rightarrow 0.5$, corresponding to identifying spikes at the intrinsic dynamical threshold $v_{th}$. 

\begin{figure}[!ht]
\begin{center}
  \includegraphics{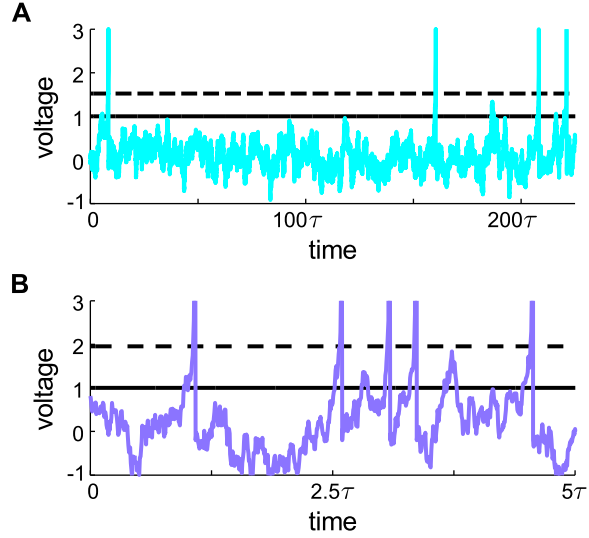}
\end{center}
\caption{ {\bf Stochastic dynamical threshold in the EIF model.}  Example voltage traces for the EIF model; voltage in units of $v_{th}-v_o$. \textbf{A}, $\sigma=0.45$. \textbf{B}, $\sigma=1.5$.  The dynamical threshold in the deterministic limit, $v_{th}=1$ in normalized units, is the solid black line.  The $95$\% confidence threshold ($C=0.95$) as defined in Eq. \eqref{StochasticThreshold2} is shown dashed.  There are multiple crossings of the deterministic dynamical threshold that do not correspond to spikes, whereas the stochastic threshold reliably identifies spikes. The stochastic threshold is \emph{adaptive} in that it depends on the input standard deviation. Some non-spiking events shown in panel B would be labeled spikes using the threshold in panel A.}
\label{EIFStochThresh}
\end{figure}

We chose $C=0.95$ to balance these competing requirements. For this value, the odds of misclassifying an aborted spike as a spike are approximately 1 in 150, and so the false-positive rate is low.\footnote{At the temporal discretization we used, and for any finite discretization, the false-positive rate is much lower than the \mbox{1 in 19} that might be expected because most spiking events do not start from exactly the threshold voltage, but from slightly larger voltages, which are more likely to correspond to true spikes.}.  We verified this procedure by computing how predictive LN models based on different choices of spike identification are, using the information criterion in Eq. \eqref{InfoLN}.  LN models that predict spikes on crossing the corresponding stochastic dynamical threshold are $1-10\%$ more informative about spiking than LN models based on spike times defined at the peak voltage (not shown), although results are qualitatively insensitive to the exact choice of threshold.  See Fig. \ref{EIFStochThresh} for example traces and discussion in context.

Armed with this treatment of spike time identification, we return to the derivation of the rate estimation function for the EIF model. Determining spike times based on the crossing of the stochastic dynamical threshold requires knowing the voltage at two adjacent times.  Eq. \eqref{PathEq} thus simplifies to:
\begin{align}
R_{\sigma}\!\left[s_x(t)\right] = \displaystyle \int \! dv(t-dt)\!\int\!dv(t)\,R[v(t-dt),v(t)]\,p_{\sigma}\!\left[v(t-dt),v(t)\big|s_x(t)\right], \label{NonLinFromVoltageEIF0}
\end{align}
where the rate based on the voltage is:
\begin{align}
R[v(t- dt),v(t)]=\frac{1}{dt}\text{H}\!\left[v_{th,\sigma}-v(t-dt)\right]\text{H}\!\left[v(t)-v_{th,\sigma}\right]; \label{RofVEIF}
\end{align}
spike times are labeled with probability one when the voltage goes from below threshold to above threshold in a single time step; this is equivalent to the condition that spike times corresponding to crossing a voltage threshold with $\dot{v}>0$ at the time of the spike. 
In Section: {\em Dynamics conditioned on the filtered stimulus}, we develop Eq. \eqref{NonLinFromVoltageEIF0} to get:
\begin{align}
R_{\sigma}\!\left[s_x(t)\right]&=\frac{\sigma\beta\!\left(C\right)}{\sqrt{2\pi\tau dt}}p_{\sigma}\!\big[v_{th,\sigma}\big|s_x(t)\big], \label{NonLinFromVoltageEIF}
\end{align}
where $\beta\!\left(C\right)$ is a constant that depends on the confidence, $C$, used to define the stochastic threshold and is defined in Eq. \eqref{Beta}.  

Eq. \eqref{NonLinFromVoltageEIF} has a simple interpretation analogous to the more intuitive result for the LIF model in Eq. \eqref{NonLinFromVoltageLIF}. The estimated firing rate is proportional to the predicted probability density at threshold, and the $\sigma$-dependent coefficient accounts for the range of voltages near threshold that are accessible in a single time step, defining an effective ``resolution'' of the threshold voltage state that depends on the interaction of the input strength and the spike-generating dynamics: 
\begin{equation}
\delta v_{th,\sigma}=\sigma\beta\!\left(C\right)\!\sqrt{\frac{dt}{2\pi\tau}}. \notag
\end{equation}
The origin of this effective threshold resolution is straightforward. First, consider $C=0.5$, for which $\beta\!\left(0.5\right)=1$ and $v_{th,\sigma}=v_{th}$, corresponding to spike times defined by crossing the intrinsic dynamical threshold.  At $v_{th}$, the voltage is described by an unbiased random walk.  The resolution, $\delta v_{th}=\sigma\!\sqrt{\frac{dt}{2\pi\tau}}$, is the mean voltage change in one time step from $v_{th}$ to voltages above $v_{th}$, and this gives the characteristic range of accessible voltages during the instant defining the spike time.  For $C>0.5$, $\beta\!\left(C\right)<1$, and so requiring increased confidence that the labeled spike time corresponds to a true spike decreases the characteristic range of voltages at the spike time---spikes are more precisely defined.    Using the effective resolution, the estimated firing rate in Eq. \eqref{NonLinFromVoltageEIF} can be be re-expressed as:
\begin{align}
R_{\sigma}\!\left[s_x(t)\right]= \frac{1}{dt}P_{\sigma}\!\left[v(t)\!\in\! \left.\left\{v_{th,\sigma}\right\} \,\right| s_x(t)\right], \label{NonLinFromVoltageEIF1}
\end{align}
where $P_{\sigma}\!\left[v(t)\!\in\! \left.\left\{v_{th,\sigma}\right\} \,\right| s_x(t)\right]$ is the predicted probability of finding the voltage in the threshold-crossing set, $\left\{v_{th,\sigma}\right\}\sim \left\{v_{th,\sigma}\lesssim v(t) \lesssim  v_{th,\sigma} +\delta v_{th,\sigma}\right\}$.
 
\begin{figure}[!ht]
\begin{center}
\includegraphics{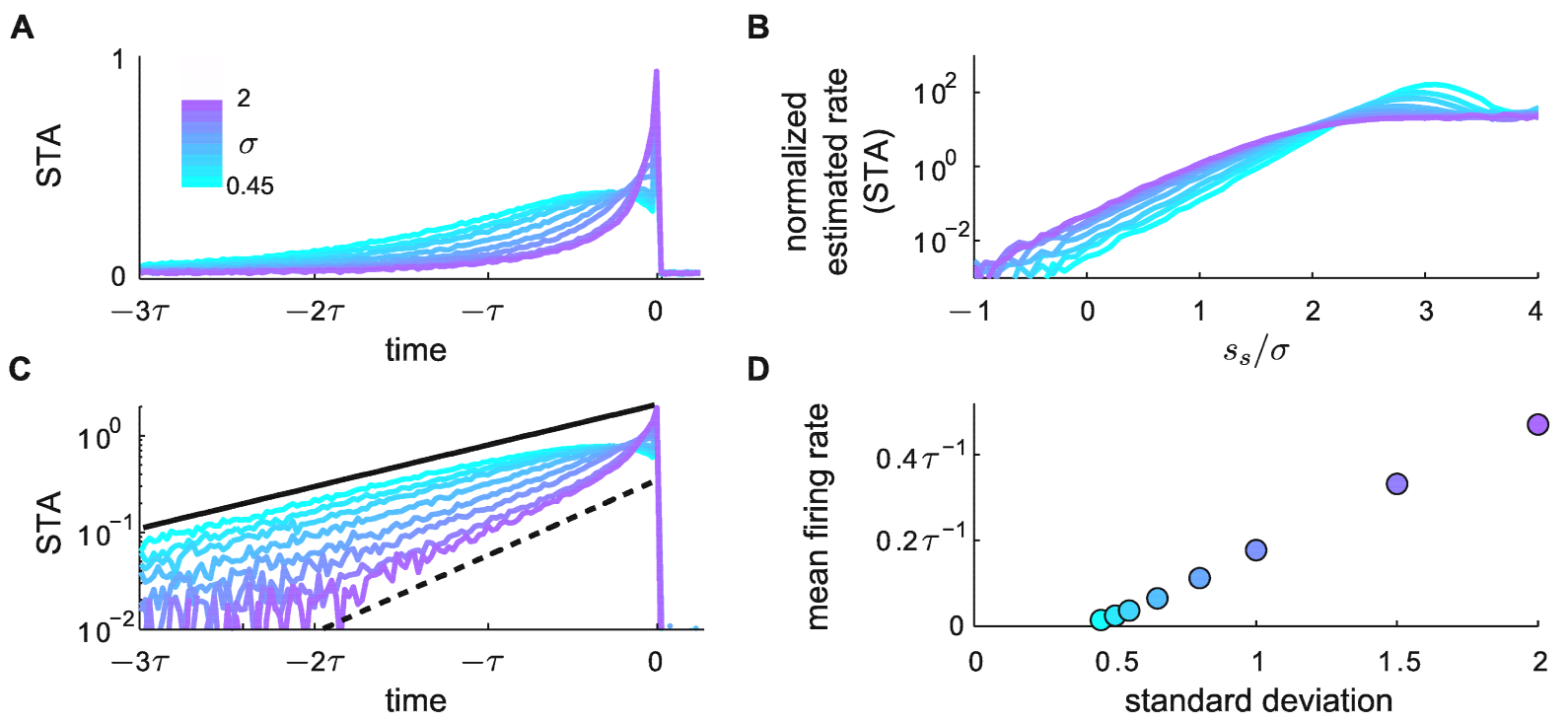}
\end{center}
\caption{ {\bf LN models for the EIF model} for $0.45\le\sigma\le 2$ using the parameters in Eqs. \eqref{DeltaGainScaling} and \eqref{VrGainScaling}.  \textbf{A} . Spike-triggered average in units of $\frac{\langle i(t)|\sp\rangle}{\sigma\sqrt{\tau/{dt}}}$. Color represents $\sigma$ and is consistent with Fig. \protect \ref{LNLIFFig}. \textbf{C}. STA on log scale.  Solid black is impulse-response filter at rest with time constant $\tau$. Dotted is the stochastic linearization filter in contrast-invariant regime, $k_{\sigma}= 1.9$ (see Eq. \eqref{KslEIF0}). \textbf{B}. Simulation results for the rate estimation function based on the STA filter, $R_{\sigma}\!\left[s_s\right]$ corresponding to Eq. \eqref{NonLinFromData}; y-axis in Log and scaled by mean firing rate; x-axis in units of relative stimulus strength, $s_s/\sigma$.    As in the case of the LIF model, estimated rate functions are graded thresholds.  For intermediate input strength, $0.8\le\sigma\le 2$, this choice of EIF parameters leads to LN models that show nearly perfect contrast gain control. \textbf{D}. Mean firing rate as a function of input standard deviation.  The EIF model parameters for contrast gain control were derived by requiring linearity in the mean rate for intermediate input strengths, $\sigma$, and over as large a range in firing rate as possible (see Section: {\em Contrast gain control in the EIF model}).
} 
\label{LNEIFFig}
\end{figure}

Results for the STA-based LN models triggered on the stochastic dynamical threshold are shown in Fig. \ref{LNEIFFig} for $0.45 \le \sigma \le 2$. For $\sigma$ larger than shown, the EIF model behavior degenerates to the LIF model behavior, and is independent of $f(v)$ provided $\Delta\ll v_s-v_{th}$.\footnote{We do not show larger inputs, $\sigma \gtrsim 2$, because they are ``unphysiological" in that input-driven fluctuations overwhelm the excitatory current, $f(v)$, and approach the amplitude of the spike itself.} 
The STA of the EIF shows more complex adaptive behavior than that of the LIF model.  At small standard deviations, the STA is non-monotonic, extends for times long compared to the membrane time constant $\tau$, and peaks well before the spike.  For intermediate input standard deviations, the STA becomes approximately exponential except at short times, with time constant close to $\tau$.   The rate estimation functions show that the EIF model is less selective to the filtered stimulus than the LIF model, as shown by the wider range of filtered stimulus values that cause spiking. 

\subsection*{The optimally predictive filter}
Our expressions for the rate estimation functions of integrate-and-fire models,  Eqs. \eqref{NonLinFromVoltageLIF} and \eqref{NonLinFromVoltageEIF}, hold for \emph{any} choice of filter. However, in order to construct the LN model that is the most precise predictor of the spike times of the dynamical model, we need to identify the most \emph{relevant} filter.  This optimal filter maximizes the information between the filtered stimulus and a spike:
\begin{equation}
h_{Opt}(t)=\operatorname*{arg\,max}_{h_x(t)} \,I^{LN}_{\sigma}[\sp;s_x], \label{HOpt}
\end{equation}
with the information given by
\begin{align}
I^{LN}[\sp;s_x]=H[s_x]-H[s_x|\sp]. \label{InfoLNG}
\end{align}
Here $H[s_x]$ is the entropy of the filtered stimulus distribution\cite{CoverThomas1991} and $H[s_x|\sp]$ is the entropy of the spike-triggered filtered stimulus distribution; equivalent forms are given in Eqs. \eqref{InfoLN0} and \eqref{InfoLN}.    The optimally predictive filter is the maximally informative dimension first introduced by Sharpee \emph{et al}\cite{Sharpee2004}.  

Given the definition of the rate estimation function in Eq. \eqref{NonLinFromVoltageEIF}, the information per spike is equivalent to the information provided by the instantaneous value of the filtered stimulus about the threshold voltage state, represented by the likelihood, $P_{\sigma}\!\left[v(t)\!\in\! \left.\left\{v_{th,\sigma}\right\} \,\right| s_x(t)\right]$. With an expression for the likelihood, the optimal filter can be derived directly from the underlying dynamics.  Unfortunately, the complexity of the expression for the conditional voltage distribution in Eq. \eqref{FormalPVS} prevents us from obtaining general analytic results.  However, we will proceed to develop intuition to guide derivations in limiting cases. We start with general considerations from information theory.  

If nothing is known about the underlying dynamics, one can assume that the STA is the optimal filter. In Eq. \eqref{InfoLNG}, the stimulus entropy, $H[s_x]$, only depends on the input $\sigma$ and is independent of the shape of the filter, and so maximizing the information minimizes the spike-triggered stimulus entropy, $H[s_x|\sp]$.  Assuming no knowledge of the dynamics and only that $p[s_x|\sp]$ is smooth and bounded, the maximum entropy (agnostic) assumption about the spike-triggered distribution is that it is Gaussian and characterized by the variance, $\text{Var}[s_x|\sp]$.  Under the maximum entropy assumption, maximizing the information is equivalent to minimizing the variance of the spike-triggered distribution. The spike-triggered variance given a spike at $t$ is:
\begin{equation}
\text{Var}[s_x|\sp]=\displaystyle \int_0^t\!\frac{dt'}{\tau}\int_0^t\!\frac{dt''}{\tau}h_x(t-t')h_x(t-t'')\left[\left\langle I(t'-t)I(t''-t)\big|\sp\right\rangle - \left\langle I(t'-t)\big|\sp\right\rangle\left\langle I(t'-t)\big|\sp\right\rangle\right]. \notag %\label{VarSSp}
\end{equation}
Since $\left\langle I(t')I(t'')\big|\sp\right\rangle $ is positive-definite for any underlying dynamics, the variance is minimized when the projection of the filter onto the STA is maximized, and thus the optimal filter is the time-reversed STA:
\begin{align}
h_{Opt}(t-t')&\propto \left\langle I(t'-t)\big|\sp\right\rangle, \notag \\
&\equiv h_s(t-t'), \label{HOptSTA}
\end{align}
where $h_s$ is the properly normalized filter. More generally, if the rate estimation function, $R_{\sigma}[s_x]$, is invertible, then, even if $p[s_x|\sp]$ is non-Gaussian, the STA is still the optimal filter. This is true because invertible transformations from filtered stimulus to estimated rate are information-preserving, and so the optimization problem reduces to the Gaussian problem above.  This \emph{minimax} solution provides one generic limit from which to study the role of the dynamics in determining the optimal filter.

A second generic limit occurs when the rate estimation function is non-invertible and is only non-zero for filtered stimulus values above some threshold: $s_x\ge s_{th}$.  If there exists a filter such that $s_{th}\gg \sigma$, then the information per spike from Eq. \eqref{InfoLN0} is dominated by the information at the stimulus threshold:
\begin{align}
I^{LN}_{\sigma}[\sp;s_x(t)]&=\displaystyle\int_{s_{th}}^{\infty}\!ds_x\,\frac{1}{\sqrt{2\pi\sigma^2}}e^{-\frac{s_x^2}{2\sigma^2}}\frac{R_{\sigma}[s_x]}{\bar{R}_{\sigma}}\log_2\!\left[\frac{R_{\sigma}[s_x]}{\bar{R}_{\sigma}}\right], \notag \\
&\approx \frac{1}{\sqrt{2\pi}}e^{-\frac{s_{th}^2}{2\sigma^2}}\frac{R_{\sigma}[s_{th}]}{\bar{R}_{\sigma}}\log_2\!\left[\frac{R_{\sigma}[s_{th}]}{\bar{R}_{\sigma}}\right], \notag
\end{align}
because the contributions from $s_x>s_{th}$ are exponentially rare in the Gaussian stimulus ensemble.  Maximizing the information in this limit is equivalent to maximizing the likelihood ratio of the voltage at threshold given the filtered stimulus at threshold; using Eq. \eqref{NonLinFromVoltageEIF} one obtains
\begin{align}
h_{Opt}(t)\rightarrow\operatorname*{arg\,max}_{h_x(t)} \,  \frac{P_{\sigma}\!\left[v\!\in\! \left.\left\{v_{th,\sigma}\right\} \,\right| s_{th}\right]}{P_{\sigma}\!\left[v\!\in\! \left\{v_{th,\sigma}\right\}\right]}. \label{HOptLowSig}
\end{align}
In this rare-events limit, the optimal filter best predicts spikes for $s_{Opt}= s_{th}$, and the quality of the predictive power of the filter for $s_{Opt}>s_{th}$ is irrelevant.  The optimal filter may differ from the STA if the dynamics that lead to spiking for filtered stimuli near threshold are distinctly different from the dynamics corresponding to spikes at larger values of the filtered stimulus, as may be the case when larger inputs cause significant inter-spike interactions\cite{Aguera2003b,Pillow2003,Powers2005}.

The details of the dynamics will determine which generic limiting optimal filter is closer to the true optimal filter.  As a proxy for understanding the information maximization problem and to develop intuition, it is useful to understand how the voltage and instantaneous firing rate correlate with the filtered stimulus.   First, we write the input in terms of the dynamics, using Eq. \eqref{LIFModel}:
\begin{align}
i(t)&=\tau\dot{v}(t)+v(t)-v_o+(v_{th}-v_r)\tau R(t), \notag\\
&=\sqrt{2}\big(\hat{h}_m^{-1}\ast (v-v_o)\big)\!(t)+(v_{th}-v_r)\tau R(t), \notag
\end{align}
where we use the inverse convolution notation as in Eq. \eqref{InverseFilterExp}. In this form, the input current is seen to drive the evolution of two correlated parts of the dynamics: the voltage and the instantaneous firing rate.  The degree of correlation between these terms depends on the timescale.  For the shortest timescales of order $dt$, the voltage and rate are strongly correlated because the dynamics are deterministic and the rate is instantaneously non-zero only at spike times.  However, since the LIF model is a renewal process and inter-spike intervals are independent, the temporal evolution of voltage decorrelates with the firing rate over multiple spikes---correlations between the voltage and firing rate on timescales comparable to the mean inter-spike interval are small. This statement is supported by the exponential decay of the spike-triggered voltage (the rate-voltage correlation function), shown in Fig. \ref{LowEigenModeFig}.
In terms of the LIF model dynamics, the filtered stimulus is:  
\begin{equation}
s_x(t)=\big(h_x\ast i\big)(t)=\sqrt{2}\left(h_x\ast \left(\hat{h}_m^{-1}\ast (v-v_o) \right) \right)\!(t)+\tau(v_{th}-v_r)\big(h_x\ast R\big)(t). \label{SofD}
\end{equation}
Eq. \eqref{SofD} is useful because it shows how the value of the filtered stimulus correlates with the dynamics.  Due to the low-pass filtering performed by $h_x$, the filtered stimulus is only correlated with the dynamics on timescales comparable to the filter duration.  Given the decaying dynamical correlations on the filter timescale,  we can approximate the voltage and rate terms as weakly correlated.

\subsubsection*{Optimal filter for the LIF model for small $\sigma$}
For small $\sigma$, the LIF model operates in the rare-events limit, $\bar{R}_{\sigma}\tau \ll 1$, and spikes are only fired in response to large deviations of the input current.  In this limit, the optimal filter satisfies the likelihood optimization criterion in Eq. \eqref{HOptLowSig}.  
For small-but-finite $\sigma$, the optimal filter is the membrane filter,
\begin{equation}
h_{Opt}=h_m, \label{HOptLowSigLIF}
\end{equation}
where $h_m$ is defined in Eq. \eqref{MembraneFilter}.  For the membrane filter, Eq. \eqref{SofD} simplifies to
\begin{equation}
s_m(t)=\sqrt{2}(v(t)-v_o)+(v_{th}-v_r)\tau\big(h_m\ast R\big)(t). \notag 
\end{equation} 
Since the term involving the rate is positive-definite, the minimum filtered stimulus that can cause spikes---the stimulus threshold---is
\begin{equation}
s_{th}=\sqrt{2}(v_{th}-v_o), \label{sth}
\end{equation}
and all spikes are fired for $s_m\ge s_{th}$.  For isolated spikes, the term involving $R(t)$ is essentially zero, the filtered stimulus is approximately perfectly correlated with the voltage trajectories leading to the spike, and so $P_{\sigma}\!\left[v\!\in\! \left.\left\{v_{th}\right\} \,\right| s_{th}\right]\approx 1$.  Thus, for $\sigma \ll s_{th}$, the membrane filter is the optimally predictive filter.  The membrane filter is guaranteed to be the optimal filter until $\sigma$ approaches $ s_{th}$, at which point the assumption leading to the likelihood optimization criterion breaks down.  

Furthermore, in this limit, the reset current after a spike is large compared to $\sigma$ and so induces a strong relative refractory period.  Because of this, the rate estimation function will have a sharp peak at $s_m=s_{th}$. Since $s_m(t)$ evolves with finite correlation time, values of $s_m(t)> s_{th}$ generally follow recently-fired spikes caused by smaller values of the filtered stimulus, and thus larger values of $s_m$ cause spikes with reduced probability.   

\begin{figure}[!ht]
\begin{center}
\includegraphics{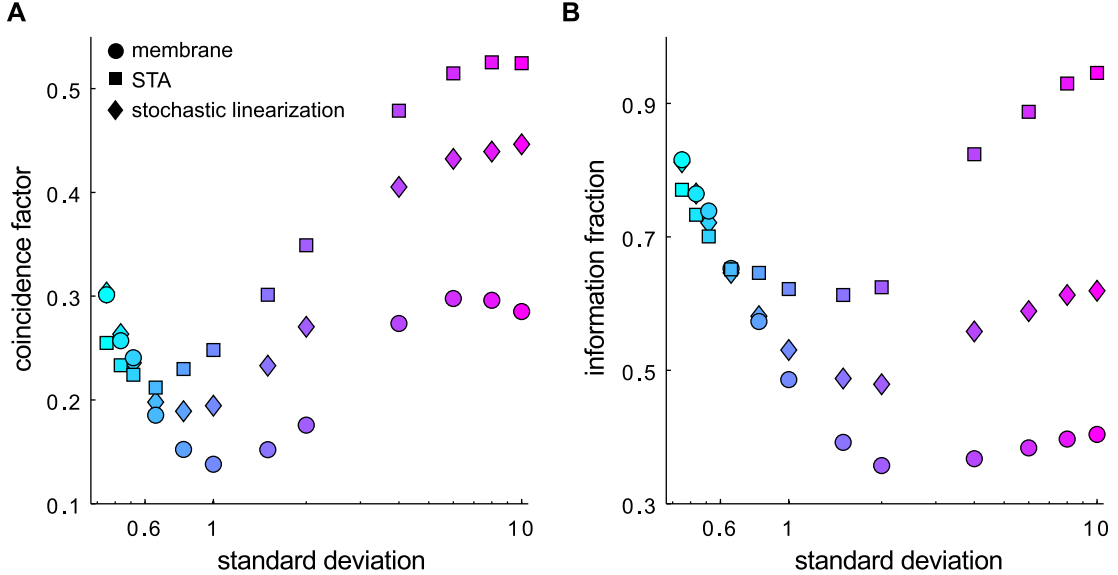}
\end{center}
\caption{ {\bf Predictive power of LN models with various filters for the LIF model.}  \textbf{A}.  Coincidence factor defined in Eq. \eqref{CoinFact} versus input standard deviation (log scale). \textbf{B}. Information per spike for each LN model relative to the information per spike in the LIF model spike train (see Eqs. \eqref{InfoD} and \eqref{InfoLN}).  Membrane filter (dots); STA filter (squares); stochastic linearization filter (diamonds).   Fig. \textbf{A} shows the coincidence factors comparing the LIF model spike trains to spike trains produced by LN models based on the intrinsic membrane filter, stochastic linearization filter, and the STA filter for temporal resolution $\gamma=\tau/20$.  At low input standard deviations, the optimally predictive filter is the membrane filter.  At large input standard deviation, the STA is the optimal filter.  The crossover occurs near $\sigma=0.6$. The stochastic linearization filter from Eq. \eqref{VoltageFilter} has been included to show how including the mean influence of spike history on the filter captures qualitatively the adaptive behavior observed empirically. Fig. \textbf{B} shows the fraction of the information captured by the LN models relative to that conferred by the LIF model spike train for temporal resolution $\tau/40$.  Trends match those of the coincidence factor, demonstrating that the fraction of information is a good measure of the predictive power of the LN models in addition to being a good measure of the completeness of the encoding representation. Absolute information per spike in the LIF model ranges from \mbox{$11.5$ bits/sp} to \mbox{$3.2$ bits/sp} for the range of $\sigma$ and temporal resolution shown. At high $\sigma$, the relative information for the STA-based LN model tends toward 1, demonstrating that the LIF model is increasingly well-described by the optimal LN model at high firing rates.}
\label{CoinFactFig}
\end{figure}

To summarize, for small-but-finite $\sigma$, one recovers as one should an LN model that is independent of the stimulus statistics and that implements the integrate-and-fire computation:  the filter is the membrane filter and spikes are predominantly fired when the filtered stimulus crosses $s_{th}$. See Fig. \ref{NonLinLowSigFig}B and accompanying details in Section: {\em Moment-based asymptotic results}.  
In this regime, the computation performed by the dynamics is dominated by the evolution of the voltage between spikes and is insensitive to inter-spike interactions. 
However, the STA is  no longer optimal for small-but-finite $\sigma$.  The STA is influenced by contributions from above-threshold stimuli that are not relevant for maximizing the information.  Except when $\sigma\rightarrow 0$ strictly, the STA differs from the membrane filter, as shown in Eq. \eqref{StaAnDt}. Accordingly, the STA-based models are less informative than the membrane filter models for small $\sigma$, as shown in Fig. \ref{CoinFactFig}B. The STA-based LN models are less precise, with lower stimulus thresholds and less selective rate estimation functions, as shown by comparison of panels B and D in Fig.  \ref{LNLIFFig}.   

\subsubsection*{Optimal filter for the LIF model for large $\sigma$}
For larger $\sigma$, the firing rate is large, $\bar{R}_{\sigma}\tau\gtrsim 1$, and inter-spike interactions play a significant role in the dynamics.  In this limit, the STA is the optimal filter:
\begin{equation}
h_{Opt}=h_s. \label{HOptHighSigLIF}
\end{equation}  
This is guaranteed to be approximately true because of the generic minimax argument leading to Eq. \eqref{HOptSTA}.  In Eq. \eqref{LimitingNonLin}, we show that the rate estimation function is approximately linearly proportional to the filtered stimulus. Thus, the spike-triggered distribution of the filtered stimulus is a skewed Gaussian, $p[s_x|\sp]\sim s_x p[s_x]$, and the optimal filter will be close to the Gaussian solution: $h_{Opt}\approx h_s$.  For $\sigma \gtrsim 0.6$, we find that the STA-based LN models have greater predictive power than models based on the membrane filter, with predictive power increasing for increasing $\sigma$, as shown in Fig. \ref{CoinFactFig}.  

For $\sigma\gg v_{th}-v_o$, we argue from the dynamics that the optimal filter is exactly the STA. 
Examine Eq. \eqref{SofD} to infer how the filtered stimulus is correlated with the dynamics \emph{for filtered stimuli that correlate with spikes}; this question corresponds to considering inputs for which the filtered instantaneous firing rate term in Eq. \eqref{SofD} is transiently (on timescales of the filter) non-zero.  First, for any filter $h_x$, for strong inputs in which the filtered firing rate term is transiently large $\big((h_m\ast R)\gg \sqrt{2}\bar{R}_{\sigma}\big)$ and thus $s_x\gg \sigma$, the value of the filtered stimulus is given primarily by the firing rate term and the voltage term is negligible. For transiently high firing rates, the voltage term is approximately fixed at $\frac{v_{th}-2v_o-v_r}{2\sqrt{2}} \ ( \ll \sigma)$: the voltage cycles between $v_r$ and $v_{th}$ and the contribution from the derivative of the voltage, $(h_x\ast\dot{v})$, goes to zero since the mean voltage is pinned.\footnote{We also use this argument to derive an asymptotic closed form for the rate estimation function, Eq. \eqref{LimitingNonLin}, which is verified in Fig. \ref{NonLinHighSigFig}A,C; see also Section: {\em LIF model: large input standard deviation}.}  Thus, considering only large values of the filtered stimulus, the STA is the optimal filter---it maximizes the correlation with the dynamics of the firing rate on timescales of the filter and is not explicitly sensitive to the details of the voltage dynamics.

Second, consider when the filtered firing rate term trends transiently to zero $\big((h_x\ast R)\ll \sqrt{2}\bar{R}_{\sigma}\big)$ and most voltage trajectories do not correspond to spikes. The minimal spiking trajectories are those that, on the timescale of the filter, evolve linearly from sufficiently far below threshold to hit $v_{th}$.  From the average minimal spiking trajectory assumed to start below rest, $\bar{v}(t)\sim \bar{v} +(v_{th}-\bar{ v})h_x(t)$ with $\bar{ v} \sim \langle v|v\le v_o\rangle=-\sigma/\sqrt{\pi}$, we infer the approximate stimulus threshold for the onset of spiking using Eq. \eqref{SofD}:  
\begin{equation}
s_{th}\sim\sigma\sqrt{\frac{2}{\pi}}, \label{SThHighSig}
\end{equation}
as is verified in Fig. \ref{NonLinHighSigFig}A.  In the previous paragraph, we observed that the STA is the optimal filter for large values of the filtered stimulus, and Eq. \eqref{SThHighSig} shows that there are no small values correlated with spiking.  Thus, for $\sigma\gg v_{th}-v_o$, the optimal filter is the STA. 
Metrics describing the predictive power of the STA-based LN model are shown in Fig. \ref{CoinFactFig}.

\subsubsection*{Filter adaptation in the LIF model}
For $\sigma\lesssim 0.6$, the optimal filter is the membrane filter and the optimal LN model is approximately independent of the input statistics.  For larger $\sigma$, the STA is the optimal filter, and furthermore, the STA adapts with changing $\sigma$, as shown in Fig. \ref{LNLIFFig}.  
The LIF model itself is fixed, and so the filter adaptation must emerge from the interaction of the input and the nonlinear dynamics. To elucidate the principles of filter adaptation, in Section: {\em Adaptation of the optimal filter}, we derive the \emph{stochastic linearization} filter, $h_l$, starting from Eq. \eqref{ErrorTerm}. The stochastic linearization filter is an approximation to the true optimal filter and accounts for the mean influence of inter-spike interactions on the optimal filter.  While it is a sub-optimal predictor of spikes for larger $\sigma$, it captures the qualitative properties of filter adaptation across all $\sigma$ and allows for easy interpretation.  

The stochastic linearization filter is the {\em exponential} filter that maximizes the correlation of the voltage and the filtered stimulus: 
\begin{equation}
h_l(t)=\sqrt{2k_{\sigma}}e^{-\frac{k_{\sigma}t}{\tau}}\text{H}(t), \label{VoltageFilter}
\end{equation}
with modified time constant:
\begin{align}
k_{\sigma}=1+\frac{ \left(v_{th}-v_{r}\right)\tau \bar{R}_{\sigma}\left(v_{th}-\langle v \rangle\right)}{\langle v^2\rangle -\langle v\rangle^2}.
\label{KslLIF}
\end{align}
The dependence of the integration time scale $k_{\sigma}$ on $\sigma$ captures filter adaptation in the LIF model by accounting for the perturbation to the linear response by the mean influence of the post-spike current.  Relative to the membrane time constant ($k=1$), the optimal filter time scale changes based on correlations of the nonlinear after-spike reset current with the voltage.  The stochastic linearization filter time scale is always shorter than the membrane time constant ($k_{\sigma}>1$) because the correction term is positive-definite.  Intuitively, the filter is shortened because spiking decorrelates the voltage and the input, leading the system to ``forget'' past inputs faster than it would by leak alone.   For the LIF model, this forgetting is the primary mechanism of filter adaptation, and it is further enhanced in the STA, where there is an additional suppression of integration at all but the shortest timescales.  More details are given in Section: {\em Semi-empirical closed form for the STA of the LIF model}.  In Fig. \ref{CoinFactFig}, we show that the stochastic linearization filter is informative across all $\sigma$ and thus quantitatively captures the phenomenology of adaptation.

\subsubsection*{Optimal filter for the EIF model}
For finite firing rates, the optimal filter for the EIF model with the parameters in Eqs. \eqref{DeltaGainScaling} and \eqref{VrGainScaling} is always approximately the STA.  This follows immediately from the general minimax argument leading to Eq. \eqref{HOptSTA}.  As shown in Fig. \ref{LNEIFFig}, the rate estimation functions are approximately exponential, and so the spike-triggered stimulus distribution is approximately Gaussian since it is the product of a Gaussian and an exponential, $p[s_x|\sp]\sim e^{s_x} p[s_x]\sim \mathcal{N}$.  Thus, for all $\sigma$ shown:
\begin{equation}
h_{Opt}\approx h_s.
\end{equation} 

As with the LIF model, the STA changes with $\sigma$ due to the interaction of the input and the nonlinear dynamics responsible for the spike.  To develop intuition, we use the trick established in Eq. \eqref{SofD} to write the filtered stimulus in terms of the dynamics:
\begin{equation}
s_x(t)=\big(h_x\ast i\big)(t)=\sqrt{2}\left(h_x\ast \left(\hat{h}_m^{-1}\ast (v-v_o) \right) \right)\!(t)+\big(h_x\ast f(v)\big)(t)+\tau(v_{th}-v_r)\big(h_x\ast R\big)(t). \label{SofDEIF}
\end{equation}
For $\sigma\ll v_{th}-v_o$, the firing rate is low and so the rate term can be ignored.  However, unlike for the LIF model, the optimal filter in the small $\sigma$ limit is not the membrane filter.  When the voltage is near the threshold state, $v_{th}$, the nonlinear spike-generating current is large compared to the input current: $f(v)\approx v_{th}-v_o\gg \sigma$.  The optimal filtered stimulus must be strongly correlated with the spike-generating current.

In previous treatments of the small-$\sigma$ limit, the STA was approximated by the most likely spike-triggered current\cite{Paninski2006STV,Wilson2008}. The most likely spike-triggered current is independent of $\sigma$ and is nonlinearly determined by the passive membrane and the form of $f(v)$.  On the approach to a spike, as $f(v)$ increases, less input current is required to drive the spike than would be the case for a purely passive membrane.  This leads to non-monotonic STAs that peak well before the spike time.  Furthermore, the excitable nonlinearity increases the influence of distant inputs preceding a spike because the nonlinearity reincorporates the input through the voltage itself. This extends the duration of the filter to timescales beyond that of the passive membrane, Fig. \ref{LNEIFFig}A, \cite{Paninski2006STV}.   

For intermediate-strength inputs, $\sigma \sim v_{th}-v_o$ and $\Delta\ll v_{th}-v_o$, the optimal filter will be close to the membrane filter.  We can argue this based on Eq. \eqref{SofDEIF}. For voltages above threshold, $f\big(v(t)\big)$ and $R(t)$ are strongly correlated and largely independent of the input. So, the optimal filtered stimulus will be most strongly correlated with the passive membrane dynamics that are primarily responsible for the approach to threshold.   The STA persists as the approximately optimal filter because the STA too becomes approximately exponential in this regime.  For the parameters studied, for intermediate $\sigma$, voltage fluctuations from the passive membrane integration are comparable to or larger than the changes in voltage caused by $f(v)$ until precisely when $f(v)$ dominates and causes a spike, independent of the input.  The STA thus primarily reflects passive integration approaching threshold.

Stochastic linearization can again be used to describe how the optimal filter adapts to changing input statistics.  For the EIF model, the modified inverse time constant is:
\begin{align}
k_{\sigma}= 1-\frac{\big\langle \left(v-\left\langle v|v\le v_{th}\right\rangle\right) f(v)\big| v\le v_{th}\big\rangle}{\text{Var}\!\left[v|v\le v_{th}\right]}+\frac{ \left(v_{th}-v_{r}\right)\tau \bar{R}_{\sigma}\left(v_{th}-\langle v |v\le v_{th}\rangle\right)}{\text{Var}\!\left[v|v\le v_{th}\right]},
\label{KslEIF0}
\end{align}
as is derived in Section {\em Adaptation of the optimal filter}.  Eq. \eqref{KslEIF0} captures the fundamental adaptive behavior of the optimal filter for the EIF model.  Again the term involving the mean rate is positive-definite and describes the more rapid "forgetting" implemented by the optimal filter due the decorrelation of voltage with past inputs due to spiking.  The new term involving subthreshold $f(v)$ reflects the influence of ``active'' subthreshold nonlinearity. For small $\sigma$, it is negative and thus acts to increase the optimal filter timescale, Fig. \ref{LNEIFFig}.  For the parameters simulated in Fig. \ref{LNEIFFig}, the subthreshold $f(v)$ term is close to zero for $\sigma \sim v_{th}-v_o$ and so forgetting dominates and $k_{\sigma}$ is always larger than $k_0$.\footnote{For $\Delta\gtrsim v_{th}-v_o$, the EIF model tends toward the quadratic integrate-and-fire model, and the subthreshold $f(v)$ term has a strong influence on the optimal filter\cite{Famulare2010}.}

\subsection*{Context-dependent coding in integrate-and-fire models}
In the previous sections, we demonstrated the principles that allow one to relate the optimal LN model to a spiking dynamical model.  Now, we switch emphasis.  We treat the LN models as the fundamental representation of the computation and discuss the properties of the integrate-and-fire models as an \emph{implementation} of the code. In this perspective, the neuron selects from its inputs the component proportional to the optimal filter.  The size of the relevant signal is proportional to the filtered stimulus, $s_{Opt}(t)$, and the remainder of the input acts as a background noise with standard deviation proportional to $\sigma$ (see Eqs. \eqref{MeanIS} and \eqref{PhiIIS}).   Because the optimal filter and corresponding rate estimation functions depend on $\sigma$, these very simple dynamical neurons effectively perform context-dependent coding: the signal that the neuron encodes and the properties of the encoding into spikes depend on the background in which the signal is embedded. 

\subsubsection*{LIF model for small $\sigma$: absolute thresholding}
As previously discussed, when the input fluctuations are small, $\sigma\lesssim 0.6$, the LN model encoding represents absolute threshold detection using the membrane filter, $h_m(t)$: the neuron spikes when the filtered stimulus crosses $s_{th}=\sqrt{2}(v_{th}-v_o)$.  In this regime, the threshold is always large, $s_{th}\gg\sigma$ and most spikes are independent; and, the optimal filter and rate estimation function are approximately independent of the input standard deviation.   This reflects the fundamental deterministic nature of the LIF model: when spikes are well-separated, the LIF model performs the integrate-and-fire computation in which the dynamics are linear integration to a fixed threshold.    

\subsubsection*{LIF model for large $\sigma$: perfect contrast gain control}
In coding terms, the LIF model performs a surprisingly sophisticated computation when the background is large. For $\sigma\gg v_{th}-v_o$, the LN models show perfect contrast gain control.    
Of all possible families of LN models, those that exhibit contrast gain control have the special property that the spike-triggered filtered stimulus distribution is independent of $\sigma$ when expressed in terms of the ratio $\frac{s_x}{\sigma}$:
\begin{equation}
p_{\sigma}\!\left[s_x(t)|\sp\right]\rightarrow p\!\left[\frac{s_x(t)}{\sigma}\bigg|\sp\right]. \label{LikelihoodGainScaling}
\end{equation}
 For example, Gaussian $p$ obeys this property as long as $\langle s_x|\sp\rangle\propto \sigma$ and $\text{Var}[s_x|\sp]\propto \sigma^2$, whereas a simple thresholding unit with threshold $s_{th}\ne 0$ and $p\left[s|\sp\right]\propto \text{H}(s-s_{th})$, does not.\footnote{Distributional scaling is defined in terms of the behavior inside an integral.  The precise meaning of Eq. \eqref{LikelihoodGainScaling} is: $\displaystyle \int\! ds\, p[s]=\int\! d\!\left(\frac{s}{\sigma}\right)p\!\left[\frac{s}{\sigma}\right] $. For example, zero-mean Gaussian $p[s]=\frac{1}{\sigma\sqrt{2\pi}}e^{-\frac{s^2}{2\sigma^2}}$ transforms as $p\!\left[\frac{s}{\sigma}\right]=\sigma p\!\left[s\right]$.}
 Comparison with Eq. \eqref{NonLinFromData} shows that the rate estimation function factors into a $\sigma$-dependent multiplicative gain given by the mean rate times a $\sigma$-independent function, $T\!\left(\frac{s_x}{\sigma}\right)$ :
\begin{equation}
R_{\sigma}\!\left[s_x(t)\right]= \bar{R}_{\sigma} \,T\!\left[\frac{s_x(t)}{\sigma}\right]. \label{NonLinCGC}
\end{equation}

\begin{figure}[!ht]
\begin{center}
\includegraphics{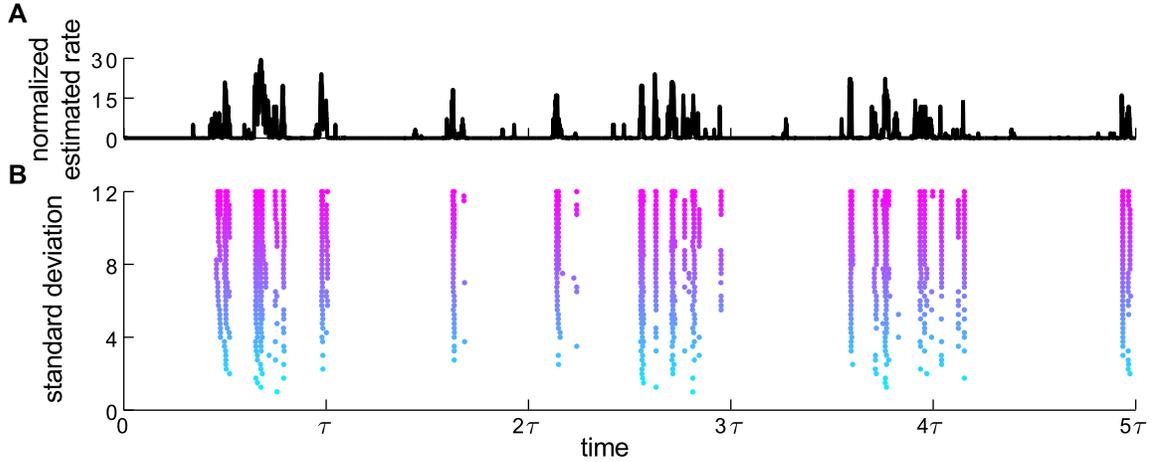}
\end{center}
\caption{ {\bf Temporal modulation of the spike times in the LIF model}. \textbf{A}. Normalized instantaneous firing rate estimate (STA-based), $R_{\sigma}\!\left[s_s(t)\right]/\bar{R}_{\sigma}$,  in the perfect contrast gain control regime ($\sigma \gtrsim 4$) for a realization of the input current.  \textbf{B}.  Spike time rasters for a fixed realization of the input current presented with increasing $\sigma$.  The spike times of the LIF models correlate with the firing rate estimate of the LN model.  Event times are reliably predicted by the LN model and are not dependent on $\sigma$. There is a $\sigma$-dependent overall increase in the mean firing rate which primarily increases the number of spikes per event.  }
\label{LIFRasters}
\end{figure}

In a system that shows this form of gain control, the thresholding on the filtered stimulus is no longer absolute. While the mean (time-averaged) firing rate, Eq. \eqref{MeanRateHighSig}, depends on the input standard deviation or typical scale, the temporal modulation of the firing rate relative to the mean does not, Fig.\ref{LIFRasters}.  One may consider this behavior to represent a form of multiplexing where different aspects of the stimulus are encoded at different observational timescales\cite{Fairhall2001,Wark2007,Panzeri2010}: individual spikes transmit context-independent information about the presence of a particular feature, while the mean rate averaged over time transmits the contextual background information. Previous work has noted aspects of contrast gain control in the LIF model\cite{Rudd1997,Paninski2003,Yu2003,Hong2008}, but we believe this is the first report of perfect contrast gain control in the LIF model. This result is surprising because the LIF model is so simple---it accomplishes this without any nonlinear processes other than the instantaneous spike generating mechanism.  

The observation of perfect contrast gain control for large $\sigma$ is confirmed with asymptotic results in Section: {\em LIF model: large input standard deviation}.  As already discussed, for large $\sigma$, the optimally predictive filter is given by the STA.  The STA is independent of $\sigma$ in the limit:
\begin{align}
h_{s}(t)\propto 3.3\sqrt{\frac{dt}{\pi\tau}}e^{\frac{k_{\infty} t}{\tau}} +\sqrt{\frac{2}{\pi}}\!\left[\left(\frac{dt}{-t}\right)^{\frac{1}{2}}-\left(\frac{5dt}{\tau}\right)^{\frac{1}{2}}\right]\text{H}\!\left[t+\frac{\tau}{5}\right], \label{LimitingSTA}
\end{align}
for $t<0$ for sufficiently small $dt$, up to normalization; this is the limiting case of Eq. \eqref{StaAnDt} with $k_{\infty}$ given in Eq. \eqref{KInfty}. The limiting rate estimation function for large $s_x$ is
\begin{align}
R_{\sigma}\!\left[s_{s}\right]\rightarrow \bar{R}_{\sigma}\left[\frac{s_{s}}{\sigma}\text{max}(h_{s})\sqrt{\pi}+\left(1-\frac{\text{max}\!\left(h_s\right)}{h_m\ast h_s}\right)\right], \label{LimitingNonLin}
\end{align}
from Eq. \eqref{NonLinHighSig}; the stimulus threshold scales with $\sigma$ and is approximately $s_{th}\approx\sigma\sqrt{\frac{2}{\pi}}$, Eq. \eqref{SThHighSig}, consistent with spiking trajectories effectively starting from the mean voltage state between spikes.  
This manifestly takes the form required for contrast gain control as shown in Eq. \eqref{NonLinCGC}: the rate estimation function is a function of the relative filtered stimulus, $s_s/\sigma$, scaled by the mean firing rate. Detailed results for the LN models are shown in Fig. \ref{NonLinHighSigFig}A.  

The absolute scale set by the threshold at $s_{th}=\sqrt{2}(v_{th}-v_o)$ seen for low $\sigma$ has vanished from the LN model. The disappearance of this intrinsic scale is a \emph{stochastically emergent} phenomenon, established only by the interaction of linear integration and instantaneous spike generation.  How does this stochastically emergent computation come about?

\subsubsection*{Principles of perfect contrast gain control in simple models}
\begin{figure}[!ht]
\begin{center}
\includegraphics{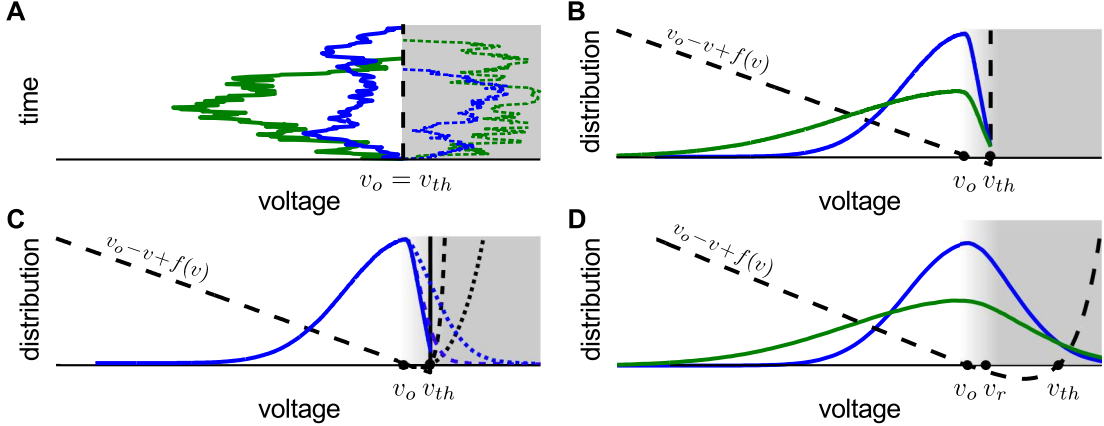}
\end{center}
\caption{ {\bf Automatic perfect contrast gain control is a generic property} of neurons that are approximately linear below threshold.  All panels: standard deviation $\sigma$ (blue) and $2\sigma$ (green).  \textbf{A}. In a neuron that is linear below threshold, voltage trajectories are realizations of the Ornstein-Uhlenbeck process between spikes\cite{GerstnerKistler}. Starting from the resting potential, typical trajectories wander from rest before returning with average displacement proportional to the input standard deviation.  The ensemble of trajectories can be broken into those that travel below rest before returning (solid) and those that travel above (dotted). When the spike-generating mechanism is included and $v_{th}-v_o\rightarrow 0$, all trajectories that would extend above threshold are killed off by the spike, but otherwise the symmetry is undisturbed and typical voltage fluctuations, which result from linearly integrating the input, remain proportional to the input standard deviation. \textbf{B}.  LIF model: steady-state voltage distributions, $p_{\sigma}[v]$ (Eq. \eqref{PVHighSig}), in the perfect contrast gain control regime for $v_r=v_o$.  Outside of the small region between rest and threshold, the distributions are half-Gaussian and scale with $\sigma$, reflecting the subthreshold linear behavior required for perfect gain control in simple neurons.  Also note the fixed point at $p[v_{th}]$. As $\frac{v_{th}-v_o}{\sigma}\rightarrow 0$, the symmetry and scaling become perfect, as predicted by Eq. \eqref{RateConstraint2}. \textbf{C}. Voltage distributions with varying $\Delta$ but fixed leak time constant and $v_r=v_o$; $\sigma$ such that the voltage distribution below rest is fixed.  When a neuron is passive below rest, in the limit of small $v_{th}-v_o$, the distributions exhibit perfect contrast gain control regardless of the details of the spike driving currents. Kinetics only affect the encoding for voltages above $v_o$, where the voltage spends little time.  \textbf{D}.  EIF model steady state distributions in the perfect contrast gain control regime for the optimal parameters in Eqs. \eqref{DeltaGainScaling} and \eqref{VrGainScaling}.  Below rest, the distributions scale with $\sigma$ as in panels A and B, again reflecting the gain control implemented via subthreshold integration.  Above rest, the exponential excitability facilitates threshold crossing, shrinking the effective distance to threshold toward zero, and effectively expanding the range of actual threshold distances that is compatible with perfect gain control.  For the optimal parameter set, the excitable dynamics facilitate gain control without otherwise disturbing the subthreshold integration, either through decreased sensitivity to the input (faster leak) or stronger inter-spike interactions (reset significantly different from rest).  
}
\label{IntuitionFig}
\end{figure}

In the large $\sigma$ limit, the LIF model shows perfect contrast gain control because it becomes statistically invariant to changes in $\sigma$.  The reason for this is evident in the complementary limit where the distances between reset and threshold, and resting and threshold, go to zero: the voltage dynamics are linear with no fixed scale.  All input integration occurs in the half-plane below threshold, and without a fixed scale, there is nothing to define a typical size of input for which the model is selective.   As shown in Eq. \eqref{PVHighSig}, the steady-state voltage distribution in terms of $\frac{v}{\sigma}$ becomes a half-Gaussian and scales linearly with contrast.    The contrast invariance is a \emph{global} phenomenon in the sense that it follows from the statistics of the voltage everywhere in the sub-threshold domain and not just at threshold.  See Fig. \ref{IntuitionFig}A,B for a depiction.

By comparing the definition of contrast gain control in Eq. \eqref{NonLinCGC} with the general form of the rate estimation function in terms of the voltage distribution in Eq. \eqref{NonLinFromVoltageEIF}, we can identify the dynamical properties that guarantee perfect contrast gain control for integrate-and-fire models.
In general, it must be true that:
\begin{equation}
\frac{\sigma\beta\!\left(C\right)}{\sqrt{2\pi\tau dt}}p_{\sigma}\!\big[v_{th,\sigma}\big|s\big]=\bar{R}_{\sigma}T\!\left(\frac{s}{\sigma}\right), \label{GainControlConstraint}
\end{equation}
for some function $T$. 
For integrate-and-fire models without any additional dynamics, the two properties that self-consistently guarantee that Eq. \eqref{GainControlConstraint} holds are
\begin{align}
\bar{R}_{\sigma}&\propto \sigma, \label{MeanRateConstraint}\\
p_{\sigma}\!\left[v\big|s\right]&\rightarrow p\!\left[\frac{v-v_{th,\sigma}}{\sigma}\bigg| \frac{s}{\sigma}\right]. \label{PVSScaling}
\end{align}
The mean rate constraint states that the rate must scale linearly to account for the $\sigma$-dependent coefficient in Eq. \eqref{GainControlConstraint}.  
The second property states that the distribution of the estimated voltage, $p[v|s]$, must depend only on the relative filtered stimulus and the scaled voltage, and that the probability of the voltage at threshold (not scaled), $p[v_{th,\sigma}|s/\sigma]$, is independent of $\sigma$. 
Averaged over $s$, Eq. \eqref{PVSScaling} implies two properties of the steady-state voltage distribution (Eq. \eqref{PVSteadyState}). First, the steady-state distribution can be expressed in terms of $v/\sigma$ only:
\begin{equation}
p_{\sigma}\!\left[v\right]\rightarrow p\!\left[\frac{v}{\sigma}\right]; \label{GlobalScaling}
\end{equation}
this is required since $s$ is a linear estimator of $v$, and so the filtered stimulus inherits its scaling properties from the voltage. Second, the probability density at the threshold is independent of $\sigma$:
\begin{equation}
 p_{\sigma}\!\left[v_{th,\sigma}\right]=\text{const}. \label{FixedThreshold}
\end{equation}
which is derived by averaging Eq. \eqref{GainControlConstraint} over $s$ when it holds.\footnote{The required invariance of the probability density at threshold to changes in $\sigma$ does not constrain the threshold to take a fixed value.  Furthermore, since contrast gain control depends on the behavior of the voltage everywhere in the subthreshold domain, its existence does not depend strongly on the precise definition of $v_{th,\sigma}$.  See Fig. \ref{IntuitionFig}D.} 
The scaling properties of the steady-state distribution imply scaling in the shifted moments:
\begin{equation}
\left\langle \left(v_{th,\sigma}-v\right)^n\right\rangle = \sigma^n \mu_n, \label{MomentScaling} 
\end{equation}
where $\mu_n$ are constants.
For the EIF model, these constraints on the voltage distribution need only hold for $v\le v_{th,\sigma}$ since the behavior of the voltage during the spike is irrelevant to the coding. 

The constraints that allow for perfect gain control in Eqs. \eqref{MeanRateConstraint}, \eqref{FixedThreshold}, and \eqref{MomentScaling} cannot always be satisfied simultaneously.  We apply our general principles to reproduce the results for the LIF model.   First, the mean firing rate is linear in $\sigma$ for large $\sigma$ as shown in Eq. \eqref{MeanRateHighSig}.  Second, when the mean firing rate is linear in $\sigma$, the probability density at threshold is fixed; using Eq. \eqref{PVSteadyStateNearThreshold},  $p_{\sigma}\!\left[v_{th}\right]\rightarrow\frac{1}{v_{th}-v_r}\sqrt{\frac{dt}{\pi\tau}}$. 
Third, the first shifted moment follows from Eq. \eqref{LIFIntegral}. Averaging over the input current gives the relationship between the mean rate and the mean voltage:
\begin{equation}
\langle v\rangle=v_o-(v_{th}-v_r)\tau\bar{R}_{\sigma}. \label{MeanRateMeanVoltage}
\end{equation}
Plugging that into the first shifted moment in Eq. \eqref{MomentScaling} constrains the mean firing rate:
\begin{equation}
\bar{R}_{\sigma}\tau=\sigma \frac{\mu_1}{v_{th}-v_r}-\frac{v_{th}-v_o}{v_{th}-v_r}. \label{RateConstraint2}
\end{equation}
This second rate constraint arrived at from the scaling behavior of the voltage distribution is only consistent with the original rate constraint in Eq. \eqref{MeanRateConstraint} when $\sigma\gg v_{th}-v_o$, independent of $v_{th}-v_r$. Thus, these constraints recapitulate the observed result: the LIF model shows perfect contrast gain control for large $\sigma$.  For mathematical subtleties and treatment of the higher order moments, see Section: {\em Higher order moment constraints for contrast gain control}.  

The mechanism for gain control is elucidated by Eq. \eqref{MeanRateMeanVoltage}: the mean voltage is set by the post-spike current, driven by the firing rate, averaged by the linear membrane dynamics. 
This gives the effective baseline state of the neuron between spikes, or ``operating point'', of the model.  Moreover, the linear subthreshold voltage dynamics guarantee that the mean rate and mean voltage are linearly related for all $\sigma$.  Thus, when the mean rate is linear with $\sigma$ and sufficiently large, the typical size of fluctuation needed to drive a spike---the distance between threshold and the mean subthreshold voltage---scales with the typical input size and thus the response is normalized. 

It is noteworthy that the spike-generating current is responsible for implementing both the slow and fast timescales in the code.  The mean rate carries the information about the statistical context, $\sigma$, while individual spikes are strongly correlated with the appearance of the feature defined by the optimal filter\cite{Fairhall2001,Wark2007,Lundstrom2008a}.  This multiplexed code can be generated with a single active current because the membrane leak is slow compared to the spike-generating current.  The mean state of the neuron between spikes is only sensitive to the mean rate, and so there is still freedom for precise modulation of the spike times on shorter time scales.

\subsubsection*{Contrast gain control in the EIF model} 
The scaling properties in Eqs. \eqref{MeanRateConstraint} and \eqref{PVSScaling} are asymptotically guaranteed to hold in the LIF model for all parameters.  However, this limit is not very biophysically plausible as the required firing rates are such that many spikes are typically fired in the integration window of the membrane and the input-induced membrane fluctuations are much larger than the intrinsic subthreshold voltage scale.  Can the more realistic EIF model also show perfect contrast gain control? 

The test of our theory is whether it allows us to \emph{predict the code} from the dynamics \emph{prior to identifying the code} with reverse correlation analysis. Accordingly, we now use the tools we have developed to predict parameter regimes for the EIF model that lead to perfect contrast gain control in the LN model representation of the encoding.  All derived results are based solely on the scaling properties of the mean dynamical response given in Eqs. \eqref{MeanRateConstraint} and \eqref{MomentScaling}; we show empirical LN models as validation.  
\newline

\noindent\emph{The limit of small distance to threshold}
\newline

In addition to the mean rate constraint in Eq. \eqref{MeanRateConstraint}, we show in Section: {\em Moment constraints for contrast gain control in the EIF model}, Eq. \eqref{RateConstraintEIF}, that subthreshold scaling with $\sigma$ in the EIF model implies a second linear rate constraint of the form in Eq. \eqref{RateConstraint2}.  Thus, both the gain control constraints are satisfied in the limit where $\sigma\gg v_{th}-v_o\rightarrow 0$, with little sensitivity to the spike generating parameters, $\Delta$ and $v_r$.  This corresponds to the limit discussed previously where there is no relevant fixed scale below threshold; see Fig. \ref{IntuitionFig}A---C.  Unlike in the LIF model, the firing rate is reasonable in this limit because the spike has finite duration. 

It is intriguing that the EIF model in this limit shows perfect gain control.  The EIF model is equivalent to the normal form of a Hodgkin Type I neuron near the bifurcation from excitable to tonic spiking\cite{Izhikevich2007}, and so qualitatively describes all Type I neurons when they are near bifurcation.  Thus, we have discovered a \emph{generic, model-independent} dynamical state that performs stochastically emergent perfect contrast gain control without any slow adaptive processes.  However, as this generic state is tied to the bifurcation point, it is finely-tuned and thus fragile to small changes in the mean of the input.  

As this limit is somewhat trivial, we will now use the constraints to find a range of parameters for the EIF model that best implement approximately perfect contrast gain control when the input is not asymptotically large.
\newline

\noindent\emph{Perfect contrast gain control with finite distance to threshold}
\newline

\begin{figure}[!ht]
\begin{center}
\includegraphics{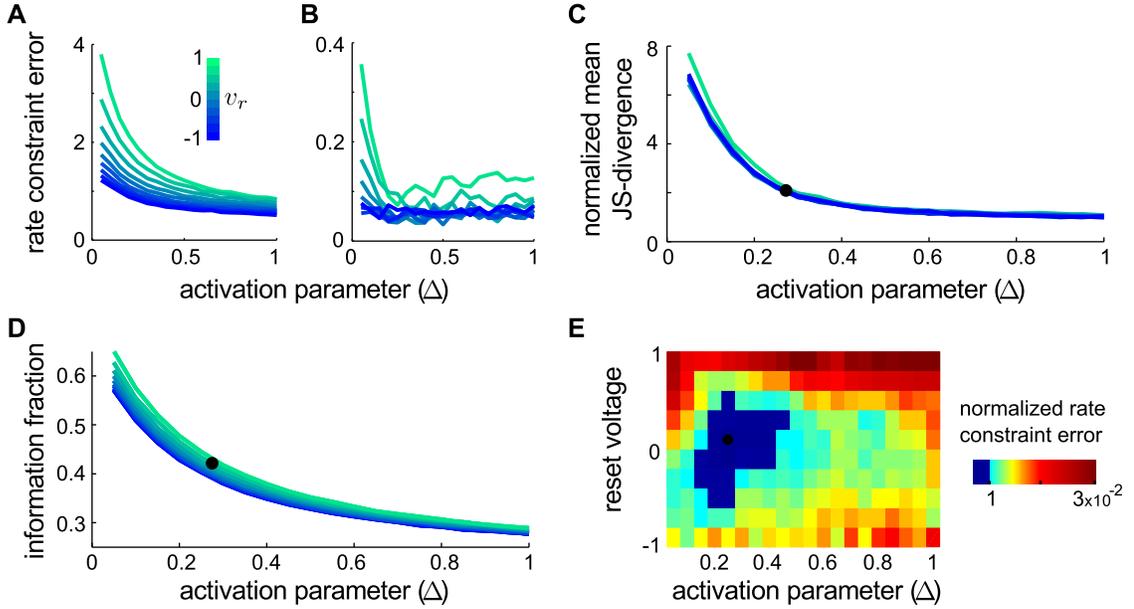}
\end{center}
\caption{ {\bf Optimal parameters for contrast gain control in the EIF model}.  All quantities are means taken over $0.5\le \sigma \le 2$.  {\bf A}. Error, Eq. \eqref{Error0}, of the rate constraint in Eq. \eqref{MeanRateConstraint}.   {\bf B}. Error, Eq. \eqref{ErrorV}, of second rate constraint in Eq. \eqref{RateConstraintEIF}.  Panels A and B together show that the dynamical constraints for perfect gain control are best satisfied for larger $\Delta$, with weak dependence on $v_r$.  
\textbf{C}.  Mean Jenson-Shannon divergence, Eq. \eqref{JSDivergence}, measuring similarity of $p_{\sigma}[s|\sp]$ across $\sigma$; y-axis normalized relative to minimum $\bar{D}_{SJ}=0.05$ bits. As predicted by the rate constraints, gain control in the LN model is best realized for larger $\Delta$.  The precision of the gain control is insensitive to $v_r$.  
 \textbf{D}.  Coding selectivity: fraction of mutual information per spike captured by the LN model relative to the deterministic spike train, as in Fig. \protect \ref{CoinFactFig}B, at temporal precision $\tau/40$. As $\Delta$ increases, the EIF model is an increasingly unreliable encoder of the filtered stimulus because of increased sensitivity to additional components of the input. 
\textbf{E}.  Optimally selective contrast gain control: mean standard deviation (Eq. \eqref{ErrorV}) of the rate constraint, normalized by the mean rate dynamic range. Minimum (black dot, also panels C,D) is the parameter set that best compromises between demonstrating selective encoding of the filtered stimulus and precise contrast gain control, Eqs. \eqref{DeltaGainScaling} and \eqref{VrGainScaling}.  Solid blue region shows parameters that do not differ significantly from the minimum. See also Fig. \protect \ref{IntuitionFig}D.
 }
\label{OptimalParameters}
\end{figure}

In Eq. \eqref{RateConstraintEIFInf}, we derive that the contrast gain control constraints are approximately self-consistently satisfied for all $\sigma$ for $\frac{\Delta}{v_{th}-v_o}\rightarrow 1$, and this result is insensitive to the reset voltage, $v_r$.  To verify this approximate analytic result, we searched over $v_r$ and $\Delta$, holding all other parameters constant, to find the models that the constraints predict will show maximally perfect contrast gain control for inputs in the range $0.5\le \sigma \le 2$.  We evaluated the deviation of the mean firing rate from best-fit linear prediction.  For constraint Eq. \eqref{MeanRateConstraint}, the error, $E^{(0)}$, is calculated with respect to the best fit with zero intercept:
\begin{equation}
E^{(0)}=\sqrt{\left\langle\left(\bar{R}_{\sigma}-\hat{R}_{\sigma}^{(0)}\right)^2\right \rangle }, \label{Error0}
\end{equation}
where $\hat{R}_{\sigma}^{(0)}$ is the best fit mean firing rate vs. $\sigma$ assuming zero intercept.  The error of the constraint from the scaling of the voltage distribution in Eq. \eqref{RateConstraintEIF}, $E^{(v)}$, is:
\begin{equation}
E^{(v)}=\sqrt{\left\langle\left(\bar{R}_{\sigma}-\hat{R}_{\sigma}^{(v)}\right)^2\right \rangle }, \label{ErrorV}
\end{equation}
where $\hat{R}_{\sigma}^{(v)}$ is the best fit mean firing rate vs. $\sigma$ with unconstrained intercept.  
Perfect contrast gain control in the LN models follows from parameter sets in the EIF model for which the above errors are minimized, with perfect gain control across all $\sigma$ occurring in the limit that $E^{(0)}$ and $E^{(v)}$ both go to zero.  Results Fig. \ref{OptimalParameters}A,B support the analytic result that the constraints are best satisfied for $\Delta\rightarrow 1$ with less sensitivity to $v_r$. 
In Fig. \ref{OptimalParameters}C, we show the mean Jenson-Shannon divergence, Eq. \eqref{JSDivergence}, to directly verify for the corresponding LN models that the precision of gain control is predicted by the precision of the constraints.
The results show that contrast gain control improves as the activation parameter, $\Delta$, gets larger.   For $\frac{\Delta}{v_{th}-v_o}\gtrsim 0.25$,
the mean divergence is within a factor of two of the minimal value for the parameter range shown.   
\newline
  
\noindent \emph{Optimally selective contrast gain control}
\newline

While contrast gain control is more precise as $\Delta$ increases, the precision comes at the cost of decreased coding effectiveness: the fraction of information per spike captured by the LN model about the filtered stimulus in the dynamical model decreases, Fig. \ref{OptimalParameters}D.  From the perspective that the dynamics are the fundamental representation of the computation, the LN model is an increasingly incomplete description of how the dynamics encode the total input current.  However, from the alternate perspective that the LN model is fundamental---that the encoding task is to select one feature as relevant and treat all other components of the input as background noise---the EIF model is an increasingly unreliable encoder of the filtered stimulus because it is increasingly sensitive to irrelevant background input.   

The decrease in coding selectivitiy with increasing $\Delta$ occurs because the larger nonlinear current, in addition to facilitating spiking, also decreases the hyperpolarized leak time constant, $\tau_L$, as shown in Eq. \eqref{TauLeak}.  The increased leakiness below the resting potential reduces the role of linear integration in spike generation and increases the sensitivity to faster components of the input relative to the STA-filtered stimulus. 
The increase in the subthreshold leak manifests itself as a reduction of the dynamic range of the response of the EIF model to the input, which in turn increases the dependence of spiking on faster components of the input.  For fixed finite $\sigma$, relative to the response of the LIF model with $\Delta=0$, the mean firing rate for finite $\Delta$ is reduced:
\begin{align}
\frac{\bar{R}_{\sigma}^{(\Delta)}}{\bar{R}_{\sigma}^{(0)}}\approx \frac{\tau_L}{\tau}, \notag %\label{RateLeakScaling}
\end{align}
as can be shown from the steady state voltage distribution in Eq. \eqref{PVSteadyState} via the re-parameterization, $v\rightarrow \sqrt{\frac{\tau_L}{\tau}}v$, for $v_r\sim v_o$.  As discussed previously after Eq. \eqref{MeanRateMeanVoltage}, the mean rate controls the mean subthreshold voltage, giving:
\begin{align}
\frac{\left\langle v|v\le v_{th,\sigma}\right\rangle^{(\Delta)}}{\left\langle v|v\le v_{th,\sigma}\right\rangle^{(0)}}\approx \frac{\tau_L}{\tau}, \notag
\end{align}
 and in the gain control regime, the mean voltage controls all higher order moments:
\begin{align}
\frac{\langle \left(v-\big\langle v|v\le v_{th,\sigma}\big\rangle\right)^n\big|v\le v_{th,\sigma}\rangle^{(\Delta)}}{\langle \left(v-\big\langle v |v\le v_{th,\sigma}\big\rangle\right)^n\big|v\le v_{th,\sigma}\rangle^{(0)}}\approx \left(\frac{\tau_L}{\tau}\right)^n, \notag% \label{MomentVLeakScaling}
\end{align}
Thus, all subthreshold moments are smaller: changes to $\sigma$ are less relevant to the dynamics for larger $\Delta$ and the voltage distribution is more narrowly concentrated with more mass near threshold, reflecting increased sensitivity to faster fluctuations in the input.

If the encoding of the filtered stimulus remained selective to a single feature in this context, increased leak would be a viable mechanism for contrast gain control in LN models.  As this is not the case, the gain control becomes increasingly trivial as $\Delta$ gets larger: while there is invariance, it is with respect to a feature to which the system is increasingly insensitive.

A selective, non-trivial contrast gain control regime is achieved by the EIF model with parameters that best compromise between implementing precise gain control while maintaining the selectiveness of the encoding by preserving subthreshold integration.  The optimal parameter set can be identified by minimizing the error in the rate constraint in \eqref{ErrorV} relative to mean rate dynamic range: $\frac{E^{(v)}}{\bar{R}_{\text{max}}-\bar{R}_{\text{min}}}$.
The optimal parameter set centered at:
\begin{align}
\frac{\Delta}{v_{th}-v_o}&=0.25, \label{DeltaGainScaling}\\
\frac{v_r-v_o}{v_{th}-v_o}&=0.1, \label{VrGainScaling}
\end{align}
as shown in Fig. \ref{OptimalParameters}E. 

\begin{figure}[!ht]
\begin{center}
\includegraphics{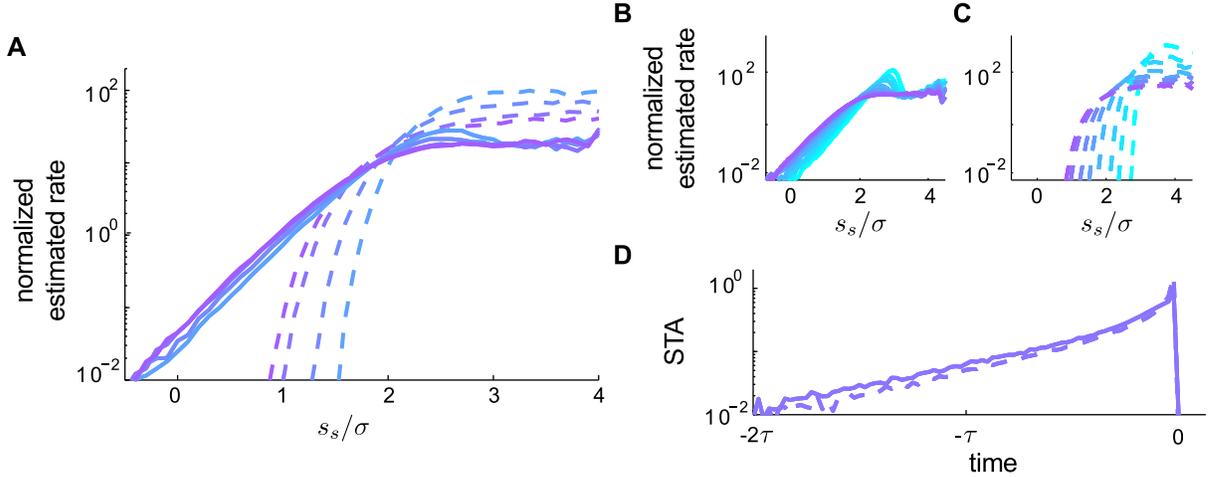}
\end{center}
\caption{ {\bf Contrast gain control comparison: EIF v. LIF models}. EIF model (solid), LIF model (dashed). \textbf{A}. STA-based rate estimation functions for each model in for $0.8\le \sigma\le 2$.  By design, the parameters used in Eqs. \eqref{DeltaGainScaling} and \eqref{VrGainScaling} lead to an EIF model that shows good contrast gain control for $\sigma\sim v_{th}-v_o$, whereas the asymptotic gain control in the LIF model (a limiting case of the EIF model) is not yet achieved.  \textbf{B}, EIF model and \textbf{C}, LIF model;   same as in panel A, but for larger input range:  $0.45\le \sigma\le 2$. \textbf{D}. STA for $\sigma=1$.  The filters are almost identical for both models.  The changes in the coding are entirely due to the differences in the spike-generating currents.  }
\label{EIFvLIFCompFig}
\end{figure}

The parameter set in Eqs. \eqref{DeltaGainScaling} and \eqref{VrGainScaling} was used to generate the LN models shown in Fig. \ref{LNEIFFig}. Comparison with the LIF model for equivalent inputs is shown in Fig. \ref{EIFvLIFCompFig}.  The rate estimation functions show precise contrast gain control for $0.8\lesssim \sigma \lesssim 2$, as opposed to for $\sigma\ge 4$ for the LIF model, a special case of the EIF model. Furthermore, the mean firing rates are reasonably low: $\bar{R}\sim 0.3/\tau$ (see Fig. \ref{LNEIFFig}D).  Some deviation from perfect contrast-invariant coding appears in the STAs in Fig. \ref{LNEIFFig}C. The long time behavior of the filters is exponential and independent of $\sigma$, agreeing with the prediction from Eq. \eqref{KslEIF0}: $k_{\sigma} \approx 1.9$. However, the filter at short times prior to the spike is sensitive to $\sigma$.  In particular, for $\sigma\lesssim 0.8$, the STA is still non-monotonic at short times, showing the influence of the subthreshold nonlinearity in $f(v)$. 

In contrast to the LIF limit ($\Delta\rightarrow 0$), in the EIF model the spike-generating currents are not perfectly time-locked, and so gain control is achieved with a less selective encoding, as shown by the broadened rate estimation functions in Fig. \ref{EIFvLIFCompFig}A and relative information in Fig. \ref{OptimalParameters}D.  The subthreshold nonlinearity also introduces a second imperfection in that the STA filter---the selected feature---is not independent of $\sigma$.  However, the deviation in the STA is due entirely to the highest frequencies in the input. In the perfect gain control regime,  LN models built on the stochastic linearization filter with $k_{\sigma} = 1.9$ capture $90\%$ of the information captured by the optimal filter at temporal resolution $\tau/40$. Furthermore, if the noise correlation time is increased to $\tau/8$, the filters become essentially pure exponentials and the rate estimation functions are qualitatively unchanged (not shown).

% Discussion can be combined with Results section as Results/Discussion
\section*{Discussion}
In this paper, we have addressed a fundamental question in computational neuroscience: how do the dynamical properties of a neuron determine its computational properties? More abstractly, what is the relationship between mechanism and meaning? The framework we have used to address this question is one in which the input is taken to be random yet fully specified; dynamics are one-dimensional; and computation is defined in terms of a "pure" coding model that does not account at all for mechanism, the linear/nonlinear model.  While we used the linear integrate-and-fire model as a stepping stone to develop our approach, our focus was on the exponential integrate-and-fire model.  Although this is still a simple model, taking it seriously is justified by its success in fitting data from certain neuronal types\cite{Fourcaud2003,Badel2008b,Badel2008c}.   This work makes two major and several minor novel contributions. First, we derive an expression for the nonlinear decision function of the resulting LN model from first principles and estimate its form in certain limits. This requires us to revisit the notion of threshold-crossing in nonlinearly excitable models and the derivation of appropriate filters for these models, issues which have some surprising subtleties. Second, we examine the requirements for a simple dynamical neuronal model to exhibit perfect contrast gain control. We find that the LIF model automatically displays this property when the input variance is large due to the loss of a typical voltage scale. Based on findings in the LIF model, we are able to show that this property also holds for certain parameter regimes in the EIF model.

The essential concept in our approach is that the filtered stimulus in the LN model is a linear estimator of the voltage in the dynamical model. Building from previous work\cite{Aguera2003a,Aguera2003b,Sharpee2004,Hong2007,Famulare2010}, we identified the optimal filter as the maximally informative instantaneous linear estimator of the threshold voltage state.  As has been previously noted, the optimal filter ``adapts'', or depends on the stimulus variance, because it has to account for the influence of the nonlinear dynamics associated with spiking on input integration in the subthreshold regime\cite{Aguera2003a,Aguera2003b,Pillow2003}.  In particular, filter timescales in simple neurons typically grow shorter as the mean firing rate of the system increases because spiking leads to forgetting: the stereotyped, input-independent spike-generating currents decorrelate the subthreshold voltage with past values of the input.  By focusing on the linear prediction error, we provided a qualitative recipe to identify the properties of the optimal filter as a function of input statistics.  Furthermore, by minimizing the mean-square prediction error below threshold, we showed how to identify the exponential tail of the optimal filter from first principles, accounting for the mean influence of inter-spike interactions on the computation performed on the input.   For the LIF model, we identified a semi-empirical closed form for the STA by combining our novel results for the long-time behavior of the filter with previous results for the short-time behavior of the spike-triggered average\cite{Paninski2006IFSTA,Badel2006}.  While we do not yet have a complete derivation, we hope this piece of progress on a long-standing question will prove useful.  

Given the filtered stimulus, we showed that the nonlinear rate estimation function is determined by the precision of the estimate of the threshold voltage state.  For models with continuous dynamics for spike generation, there is no unique threshold voltage state.  For an example, the EIF model, we provided a recipe to identify an appropriate choice of the threshold voltage state---the stochastic dynamical threshold---that captures how the choice to assign a unique spike time to a continuous spike waveform depends on the spike-generating dynamics and the input statistics.  Our definition generalizes the concept of an intrinsic dynamical threshold found in previous work\cite{Izhikevich2007,GerstnerKistler,Hong2007,Famulare2010} to better represent spike initiation behavior with finite noise strength.  LN models are imperfect predictors of the spike train of the dynamical model because the linear estimate is necessarily imperfect. 

To study the voltage estimation problem analytically, we introduced the conditional dynamical process that is completely specified by $p\!\left[v(t')\big|v(t'-dt),s(t)\right]$ in Eq. \eqref{ConditionalProcess}. It is a time-inhomogeneous (non-stationary) Markov process for fixed $t$, and the general two-parameter process for flowing $t$ is both non-stationary and non-Markov.  This turns out to be a specific example of a conditional Markov process, first introduced by Stratonovich\cite{Stratonovich1960,Stratonovich1968}.
To construct the conditional dynamical process, we studied the input conditioned on observing the filtered stimulus, $p\!\left[i(t')\big|s(t)\right]$. 
The conditional input distribution is a precise representation of the concept of feature selection.  The input, $i(t)$, is decomposed into an observed component, $s(t)$, and a background noise source whose strength is characterized by the input standard deviation, $\sigma$.  Because we use the dynamics to identify the relevant component that is observed by the neuron, this \emph{post hoc} separation of source and signal can be thought of as ``taking the neuron's perspective'' where it must automatically separate background from signal without any \emph{a priori} knowledge.   

In contrast, the recent work on the derivation of LN models from integrate-and-fire models by Ostojic and Brunel\cite{Ostojic2011} assumes that the signal is \emph{a priori} separable from the noise background, as has been the case for all previous analytic work on reducing spiking models to rate models known to us (most notably also including Plesser \emph{et al}\cite{Plesser2000}).  Our formulation is equivalent to the approach of Ostojic and Brunel if we break the correlation of the signal and the noise in our conditional input distribution, and instead take the conditional input distribution to be characterized by the moments $\langle i(t')|s(t')\rangle = s(t')$ and $\phi_{ii|s}(t',t'')=\sigma^2\tau\delta\!\left(t'-t''\right)$ in place of Eqs. \eqref{MeanIS} and \eqref{PhiIIS}.  Under this assumption about the \emph{a priori} independence of signal and noise, the conditional dynamical process remains Markov, which enables those authors to use elegant Fokker-Planck methods to derive the filter and the rate estimation function in place of our more cumbersome voltage state-space approach.  

Given white noise inputs, we showed that the LIF model shows two distinct computational regimes depending on the statistical context. In the small and large background limits, the LN model is predictive of the dynamical response and yet the codes are fundamentally different.  This is an example of context-dependent coding.  In the small background limit, the LIF model integrates the input with a fixed filter and fires spikes based on an absolute threshold, while for large background, the LIF model uses single spikes to encode relative fluctuations of a feature represented by a different filter.  Thus, the simplest possible intrinsic neural dynamics interact with the input to perform sophisticated adaptive computation.  Furthermore, if we were to present an input with slowly changing standard deviation, the adaptation of the code happens on the timescale of a single inter-spike interval because these models are renewal processes.  In other words, complete adaptation occurs after a single spike.  This work may prove to be useful for understanding very rapid adaptive changes observed in biological systems, including the fly H1 neuron\cite{Fairhall2001} and the retina\cite{Kim2001,Baccus2002}.  

While previous work has shown gain control in the LIF model\cite{Rudd1997,Paninski2003,Yu2003}, this is the first study to show the limit of perfect gain control.  By studying $p\!\left[v(t)\big|s(t)\right]$, we identified the general principles behind this adaptive computation and demonstrated that this type of gain control is generic to Type 1 neurons whenever the neuron is close to bifurcation.  Furthermore, we were able to use the dynamical constraints associated with perfect contrast gain control to derive optimal parameters for the EIF model to show approximately perfect gain control for finite distance between rest and threshold.  To our knowledge, this is the first time theory has been used to \emph{design} from first principles a dynamical neuronal model to implement a specific type of code. 

The concepts we used to derive LN models from dynamical models directly generalize to more biophysically plausible dynamical neuron models. First, the EIF model has been shown to be a useful general reduction of more complex biophysical models and can reproduce cortical cell recordings\cite{Badel2008b,Badel2008c} and so our results will be directly applicable when that is the case.  However, more generally, neuronal systems are described by multidimensional models. Our approach to the voltage estimation problem can be extended to include considerations of the states of ion channels. The stochastic dynamical threshold defined similarly will generically be multi-dimensional, and the space of relevant filters may be multidimensional as well\cite{Azouz2000,Aguera2003a,Hong2007}.   Explicit spike-history dependence can also be included in the conditional dynamics framework as an additional spike-dependent filter, as in generalized linear models\cite{Paninski2004}.

The concept of the conditional dynamical process may prove more generally useful for studying the dynamical implementation of neural computation.  Important goals of research in neural computation are to identify the nervous system's abstract representations of relevant stimuli, internal states and behaviors, and to discover the biological mechanisms used to implement and manipulate these abstractions.  Here, we used the conditional dynamical process to associate an abstract quantity, the filtered stimulus, with sub-ensembles of states of a dynamical system, the sets of voltage trajectories that cross threshold and are consistent with a given value of the filtered stimulus.  
This tool provided the bridge that enabled us to use methods from either coding or dynamics to simultaneously study both.  
While the problem studied here is often analytically tractable because the underlying unconditional dynamics are Markov, the dimensionality is low, and feature selection is linear, these details are not essential to the concept.  As similar ideas were first introduced by Stratonovich in the contexts of state estimation and nonlinear control\cite{Stratonovich1960,Stratonovich1968}, we are optimistic that this work will lead to a program to import existing techniques from those areas that can be used to study the biophysics of neural coding and to deepen the connections between neural coding and control.

Beyond computational neuroscience, the conditional process derived here provides what appears to be a novel approach to studying the accuracy of the stochastic linearization approximation to a nonlinear system.  Stochastic (or statistical) linearization has primarily found applications in nonlinear control and the study of engineered structures subject to random vibrations such as earthquakes and ocean waves\cite{Socha2005p2}.  An underdeveloped aspect of the theory of stochastic linearization is error analysis of the linear approximation\cite{RobertsSpanos,Socha2008}.  The conditional distribution, $p\big[v(t)\big|s(t)\big]$, is the distribution of the true state of the system given the linear approximation, and thus it captures completely the error of the approximation locally in state space.  This has the potential to be a more precise tool than the mean-square error and steady-state distribution-based metrics that are commonly used.  Such state-dependent error estimation can be important when tolerances or failure modes are poorly described by global measures of error\cite{Naess1992,Naess1995}.  We hope that this work may seed further development in this domain.

% You may title this section "Methods" or "Models". 
% "Models" is not a valid title for PLoS ONE authors. However, PLoS ONE
% authors may use "Analysis" 
% can be moved in front of results/discussion

\section*{Models and Methods}
\subsection*{Integrate-and-fire models}
We work with the integrate-and-fire models, and we restrict ourselves to excitable cases where model exhibits a stable rest state in the absence of input.  We parameterize the models as:
\begin{align}
\tau \dot{v}&=-v+v_o+f(v)-\left(v_{s}-v_{r}\right)\tau R(t)+ i(t), \notag \\
f(v)&= \left(v_{th}-v_o\right)\left(e^{\frac{v-v_{th}}{\Delta}}-\left(1+\frac{v-v_o}{\Delta}\right)e^{\frac{v_o-v_{th}}{\Delta}}\right) \left(1-\left(1+\frac{v_{th}-v_o}{\Delta}\right)e^{\frac{v_o-v_{th}}{\Delta}}\right)^{-1},  \label{LIFModel}\\
R(t)&=\delta\!\left(v-v_{s}\right) \dot{v}\text{H}(\dot{v}), \notag %\label{LIFRofT}
\end{align}
where $v_o$ is the resting potential for zero input, $v_r$ is the reset voltage immediately after a spike, $\tau$ is the ``physiological'' membrane time constant near rest, $v_s$ is the peak voltage of the spike and the input current is in units of voltage (the input resistance is set to one) for convenience.  The function $f(v)$ is the exponential voltage-activated spike generating current, where $\Delta$ sets the activation scale over which the spike-driving excitable current turns on.  There is an unstable fixed point at $v_{th}$ (for $v_{th}>v_o$) that acts as the intrinsic \emph{dynamical threshold} for pulse-like inputs\cite{Azouz2000,Izhikevich2001,GerstnerKistler,Izhikevich2007,Hong2007,Famulare2010}. This form of $f(v)$ has been chosen so that the resting potential, threshold, and membrane time constant at rest are independent of other parameters.  The leak time constant, $\tau_L$, for hyperpolarized voltages, $v\ll v_o-\Delta$, is: 
\begin{align}
\tau_L&=\tau\left(1-\frac{\frac{v_{th}-v_o}{\Delta}}{\left(e^{\frac{v_{th}-v_{o}}{\Delta}}-1\right)}\right), \label{TauLeak}\\
&\!\begin{array}{lc}
\rightarrow \tau & \text{for}\quad \frac{\Delta}{v_{th}-v_o}\ll 1,\\
\\
\rightarrow \frac{\tau}{2}\left(\frac{v_{th}-v_o}{\Delta}\right) & \text{for}\quad \frac{\Delta}{v_{th}-v_o}\gg 1. 
\end{array}\notag
\end{align}
We describe the dynamics of the after-spike reset to $v_r$ with the deterministic instantaneous firing rate, $R(t)$,  given above.  The continuous-time, voltage-based reset operation,  $\left(v_r-v_{s}\right)\tau \,\delta\!\left(v-v_{s}\right) \!\dot{v}\text{H}(\dot{v})$, is read as ``when the voltage reaches the spike height $v_{s}$ from below at time $t$, reset the voltage to $v_r$.''\cite{Paninski2003}

Of the model's six parameters, only two are meaningful from the perspective of the geometry of the dynamics; the others determine units and the finite cut-off for the spike height (to which the model is quite insensitive\cite{Touboul2009b}).  We treat $\Delta$ and $v_r$ as the meaningful free parameters. 

For finite $\Delta$, Eq. \eqref{LIFModel} is the exponential integrate-and-fire (EIF) model\cite{Fourcaud2003}.  In the limit of $\Delta=0$, we recover the leaky integrate-and-fire (LIF) model.  Considered this way, the LIF model is linear below $v_{th}$, and whenever the voltage exceeds $v_{th}$, it instantaneously jumps to $v_s$ and back down to $v_r$ (where instantaneously means faster than the shortest timescale in the stimulus or subthreshold dynamics). The usual definition of the LIF model with $f(v)=0$ and $v_s=v_{th}$ is equivalent in probability since the time spent in the interval between $v_{th}$ and $v_s$ is zero. In the limit of $\Delta\rightarrow \infty$, the model becomes the quadratic integrate-and-fire model\cite{Ermentrout1995}; we do not attend to this limit in this paper.

In the figures, all results are displayed in terms of the intrinsic scales, $v_{th}-v_o$ and $\tau$.   In LIF simulations, we used $v_r=v_o$, and no results presented here qualitatively depend on this choice.  For the EIF, in simulations, we used the parameters in Eqs. \eqref{DeltaGainScaling} and \eqref{VrGainScaling}, and $v_s-v_o=20\left(v_{th}-v_o\right)$.  

It is useful to view the model in integral form:
\begin{align}
v(t)=v_o+\displaystyle \int_0^{t}\frac{dt'}{\tau}e^{\frac{t'-t}{\tau}}\left[i(t')+f\big(v(t')\big)-\left(v_{s}-v_{r}\right)\tau R(t')\right], \label{LIFIntegral}
\end{align}
ignoring any initial transient.  Thus, the voltage is determined by the action of an exponential filter on the input current and the nonlinear currents due to spiking at times prior to $t$.  The filter is determined by the linear membrane dynamics near rest and so we refer to it as the intrinsic \emph{membrane filter}:
\begin{equation}
h_m(t)=\sqrt{2}e^{-\frac{t}{\tau}}\text{H}\!\left[t\right], \label{MembraneFilter}
\end{equation}
where the $\sqrt{2}$ comes from our normalization convention for filters ($\int\!\frac{dt'}{\tau}h(t')^2=1$) and the Heaviside function accounts for causality.

We examine how a spiking dynamical neuron encodes a realization of a stimulus drawn from a given stimulus ensemble.
We will work with Gaussian white noise input currents with fixed mean and autocorrelation. Such stimuli richly explore the computational properties of the neuron \cite{RiekeSpikes}.  While the elegant mathematical properties of such a stimulus is helpful, it may also be a reasonable simplified model of synaptic inputs in cortex\cite{Destexhe1999}.  We consider the dynamics to be a deterministic response to a known realization of white noise.  We will consider input ensembles with zero mean and autocorrelation function, $\phi_{ii}$:
\begin{align}
\phi_{ii}\!\left(t-t'\right)\equiv \langle i(t)i(t')\rangle-\langle i(t)\rangle\langle i(t')\rangle=\sigma^2 \tau\delta\!\left(t-t'\right),  \notag %\label{Phiii}
\end{align}
where $\sigma$ characterizes the typical scale of fluctuations.

For integrate-and-fire models driven by white noise, the steady state voltage distribution and the mean firing rate can be calculated directly from the Fokker-Planck formulation of the model\cite{Risken}.  Previous work \cite{Paninski2003,Fourcaud2003} has shown that:
\begin{align}
p_{\sigma}\!\left[v\right]=\frac{2\bar{R}_{\sigma}\tau}{\sigma^2}\displaystyle  e^{-\frac{(v-v_o)^2-2F(v)}{\sigma^2}}\int_{\text{max}(v,v_r)}^{v_{s}}dv'\,e^{\frac{(v'-v_o)^2-2F(v')}{\sigma^2}},\label{PVSteadyState}
\end{align}
where $F(v)=\int\!dv\,f(v)$ and the mean rate, $\bar{R}_\sigma$ is found from the normalization condition: $\int \!dv\, p_{\sigma}\!\left[v\right]=1$.  

For the LIF model, in the limit where the input standard deviation is large compared to $v_{th}-v_o$ and $v_{th}-v_r$, the mean rate is:
\begin{equation}
\bar{R}_{\sigma\rightarrow \infty}=\frac{\sigma}{\sqrt{\pi}\left(v_{th}-v_r\right)\tau}, \label{MeanRateHighSig}
\end{equation}
and the steady state voltage distribution is:
\begin{equation}
p_{\sigma\rightarrow \infty}\!\left[v\right]=\left\{
         \begin{array}{ll}
                   \frac{2}{\sqrt{\pi}\sigma}e^{-\frac{v^2-v_r^2}{\sigma^2}} & \text{if } v\le v_r, \\
                  \frac{2}{\sqrt{\pi}\sigma} \frac{v_{th}-v}{v_{th}-v_r} & \text{if } v_r< v<v_{th},
        \end{array}
     \right. \label{PVHighSig}
\end{equation}
as can be derived by manipulating Eq. \eqref{PVSteadyState} with $\int_x^{x+\epsilon}f(y)dy\approx \epsilon f(x)$ in mind.

\subsubsection*{Discretization and regularization}
In this work, we use both continuous and discrete time notations, depending on which is more natural.  All continuous time expressions should be interpreted in the Ito sense, and all discrete time expressions are given by the corresponding Euler-forward discretization of the continuous system\cite{Risken}.   Explicitly, in discrete time, the models become:
\begin{align}
v(t)=\left\{ 
           \begin{array}{lr} 
                            v(t-dt)+\frac{dt}{\tau}\!\left[-v(t-dt)+v_o+f\big(v(t-dt)\big)+i(t-dt)\right]  & \text{if } v(t-dt)< v_{s} ,\\
                            v_r +\frac{dt}{\tau}\!\left[-v_r+v_o+f\big(v_r\big)+i(t-dt)\right]  & \text{if } v(t-dt)\ge v_{s},
           \end{array}
       \right.  \label{LIFModelDiscrete}         
\end{align}
where we have defined the reset to be non-anticipating. In this discretization, spikes are reset in the EIF model whenever $v(t)\ge v_s$, and spikes occur in the LIF model whenever $v(t)\ge v_{th}$ (since again, for $\Delta=0$, any voltage above $v_{th}$ is instantaneously sent above $v_s$).  For clarity of exposition, we find it natural to describe the input in terms of the physical current $i(t)$.    To preserve the white noise statistics of the input, the discretization of the current is: $i(t)=\sigma\sqrt{\frac{\tau}{dt}}\xi(t)$, where $\xi(t)$ is discrete Brownian motion with zero mean and unit variance.  The input current strength is thus a function of the discretization time scale, and, in the continuous limit, diverges.  This is the usual pathology that often arises when taking white noise seriously in a physical setting.  

The instantaneous discontinuity at threshold in the LIF model causes an additional pathology in the continuous limit: the spike-triggered average input (STA) current (Eq. \eqref{STA}), is a singular function of $dt$ and diverges \cite{Paninski2006IFSTA,Badel2006}.   For finite $dt$, this manifests as a boundary-crossing contribution to the STA that exists at short times prior to the spike and is proportional to $\sigma$ in amplitude\cite{Aguera2003b,Paninski2006IFSTA,Badel2006}.  We study the STA in more detail in the Section: {\em Semi-empirical closed form for the STA of the LIF model}.  This boundary effect also exists in the EIF model, but it is generally negligible because the approach to the reset at short times is dominated by the intrinsic dynamics of the exponential current and the input plays essentially no role.

A simpler divergence also exists in the STA of the EIF model when triggered on the stochastic dynamical threshold defined in Eq. \eqref{StochasticThreshold2}.  With true white noise inputs, the condition that threshold is crossed from below biases the average input current in the last sample preceding the spike to be positive.  This leads to a ``delta-function'' component in the STA immediately preceding the spike\cite{Aguera2003b,Hong2007,Famulare2010}. This threshold-crossing component only has support over one time bin and appears generically in any continuous dynamical model driven by white noise.

In a physical neuron, the pathologies associated with true white noise are not relevant because inputs have finite correlation time and true spiking mechanisms are not infinitely fast.  Thus, throughout this work, there is an implicit regularization: $dt$ should be thought of as the small-but-finite correlation time of the input.  In expressions that are sensitive to this regularization, $dt$ appears explicitly; otherwise, $dt$ naturally drops out in the limit. All time series are displayed with resolution $dt=\frac{\tau}{40}$; at larger variances, the simulations were run with smaller hidden time step and downsampled for presentation.
  
The discrete time model can be expressed as a transition probability. Eq. \eqref{LIFModelDiscrete} is equivalent to the transition probability density:
\begin{align}
p\big[v(t)\big|& v(t-dt),i(t-dt)\big]= \notag\\
&\delta\!\left[v(t)-v(t-dt)-\frac{dt}{\tau}\!\left[-v(t-dt)+v_o+f\big(v(t-dt)\big)+i(t-dt)\right]\right]\text{H}\!\big[v_{th}-v(t-dt)\big]  \label{LIFPropagator}\\
&\qquad +\delta\!\left[v(t)-v_r-\frac{dt}{\tau}\!\left[-v_r+v_o+f\left(v_r\right)+i(t-dt)\right]\right]\text{H}\!\big[v(t-dt)- v_{th}\big].\notag
\end{align}
Marginalizing over the input current gives:
\begin{align}
p\big[v(t)\big| v(t-dt)\big]=&\frac{\text{H}\big[v_{th}-v(t-dt)\big]}{\sqrt{2\pi\sigma^2\frac{dt}{\tau}}}e^{-\frac{\left(v(t)-v(t-dt)-\frac{dt}{\tau}\!\left[-v(t-dt)+v_o+f\left(v(t-dt)\right)\right]\right)^2}{2\sigma^2 \frac{dt}{\tau}} } \notag \\
&\quad +\quad\frac{\text{H}\big[v(t-dt) -v_{th}\big]}{\sqrt{2\pi\sigma^2\frac{dt}{\tau}}}e^{-\frac{\left(v(t)-v_r-\frac{dt}{\tau}\!\left[-v_r+v_o+f\left(v_r\right)\right]\right)^2}{2\sigma^2 \frac{dt}{\tau}} } , \label{FreeProcess}
\end{align}
which describes the dynamics of the model if we do not observe the input current at time $t-dt$.

For the LIF model in discrete time, to order $\sqrt{dt/\tau}$, the steady state voltage distribution in discrete time is unchanged from the continuous-time result in Eq. \eqref{PVSteadyState} except for voltages near $v_{th}$.   The leading correction to the steady state distribution for voltages near threshold, $v\sim v_{th}\pm \sigma\sqrt{\frac{dt}{\tau}}$, can be derived by propagating the continuous time steady-state distribution forward one time step near threshold for the typical range of voltages that can be spanned in one time step:
\begin{align}
p_{\sigma}\!\left[v|v\ge v_{th}\right]&\approx \displaystyle \int_{v_{th}-\sigma\sqrt{\frac{dt}{\tau}}}^{v_{th}} dv' p\big[v\big| v'\big]p_{\sigma}[v], \notag \\
 &\approx\frac{\bar{R}_{\sigma}dt}{\sqrt{\pi\sigma^2\frac{dt}{2\tau}}}e^{-\frac{\left(v-v_{th}\right)^2}{2\sigma^2\frac{dt}{\tau}}}. \label{PVSteadyStateNearThreshold}
\end{align}

\subsection*{Identifying LN models with reverse correlation}
All analysis based on simulation data used reverse correlation to find the linear-nonlinear models corresponding to each stimulus condition\cite{deBoer1968,Hunter1986,Sakai1992,RiekeSpikes}. 
The standard choice for the filter is the spike-triggered average current (STA).  To find the STA, we average the input current preceding each spike:
\begin{equation}
\langle i(t)|\sp\rangle=\frac{1}{N}\displaystyle \sum_{i=1}^Ni(t-t_i), \label{STA}
\end{equation}
where the $\{t_i\}$ are the times of the spike.  For the LIF model, spikes times correspond to the instants when the voltage exceeds $v_{th}$.  For the EIF model, we identify spike times as the instants for which the voltage crosses the stochastic dynamical threshold defined in Eq. \eqref{StochasticThreshold2} from below. 

Given a choice of filter, $h_x$ (where the subscript provides a label in context), we construct the filtered stimulus, $s_x(t)$, by convolving the input current with the filter:
\begin{equation}
s_x(t)=\displaystyle \int_0^t \frac{dt'}{\tau}h_x(t-t')i(t')=\big(h_x \ast i\big)(t).
\label{s_x}
\end{equation}
Our filters are causal and so take the form of a continuous function multiplied by the Heaviside step function,  $\text{H}(t-t')$. Consistency with Ito calculus requires $H(0)=0$\cite{Risken} . We normalize filters such that $\int\!\frac{dt'}{\tau}h(t')^2=1$. With this choice, the variance of the filtered stimulus is always: $\langle s_x(t)^2\rangle=\sigma^2$.  

With the filtered stimulus,  the distribution of filtered stimuli given a spike,
\begin{equation}
p_{\sigma}\!\left[s_x(t)|\sp\right], \notag %\label{PsSp}
\end{equation}
can be sampled with reverse correlation\cite{deBoer1968}. The rate estimation function for a particular filter,  $R_{\sigma}\!\left[s_x(t)\right]$, can then be found via Bayes rule:
\begin{align}
R_{\sigma}\!\left[s_x(t)\right]&\equiv \frac{P_{\sigma}\!\left[\sp|s_x(t)\right]}{dt} \notag\\
&=\bar{R}_{\sigma}\frac{p_{\sigma}\!\left[s_x(t)|\sp\right]}{p_{\sigma}\!\left[s_x(t)\right]}, \label{NonLinFromData}
\end{align}
where the prior distribution of stimuli, $p_{\sigma}\!\left[s_x(t)\right]$ is always Gaussian with mean zero and variance $\sigma^2$.  By choosing the normalization of the filter as described above, all changes in gain appear in the shape of $R_{\sigma}\!\left[s_x(t)\right]$.  The filter subscript emphasizes that the threshold function and the spike-triggered distribution generally depend strongly on the input standard deviation and the choice of filter.  Along with the changes in the optimally predictive filter, this dependence is a form of adaptive coding that we will elucidate here.

\subsection*{Quantifying the precision of contrast gain control}
In Fig. \ref{OptimalParameters}C, we quantify the precision of the contrast gain control with the Jenson-Shannon divergence\cite{CoverThomas1991} of the spike-triggered stimulus distribution at $\sigma$ in terms of $z\equiv \frac{s_x}{\sigma}$ relative to the distribution at $\sigma=1$, averaged over $\sigma$,
\begin{equation}
\bar{D}_{SJ}=\frac{1}{2}\displaystyle\int\!dz\!\left\langle p_{\sigma}[z|\sp]\log_2\!\left[\frac{p_{\sigma}[z|\sp]}{m_{\sigma}[z|\sp]}\right]+p_{1}[z|\sp]\log_2\!\left[\frac{p_{1}[z|\sp]}{m_{\sigma}[z|\sp]}\right]\right\rangle_{\!\!\sigma}, \label{JSDivergence}
\end{equation}
where $m_{\sigma}[z|\sp]=\frac{1}{2}\left(p_{\sigma}[z|\sp]+p_1[z|\sp]\right)$. When $\bar{D}_{SJ}\rightarrow 0$, contrast gain control is perfect---the scaling relations in Eqs. \eqref{LikelihoodGainScaling} and \eqref{NonLinCGC} hold for all $\sigma$.  Results are insensitive to this choice of metric.

\subsection*{Coincidence factor}
With respect to the  encoding used in the decision to fire a spike, the best LN model description of the dynamical model optimally predicts the spikes of the deterministic system.  To measure the predictive power directly, we calculate the coincidence factor from artificial spike trains generated from an inhomogeneous Poisson process with the instantaneous rate given by the LN model. The coincidence factor is:
\begin{equation}
\Gamma=\frac{N_{coinc}-\langle N_{coinc}\rangle}{\frac{1}{2}\left(N_{data}+N_{model}\right)}\mathcal{N}^{-1}, \label{CoinFact}
\end{equation}
where $N_{coinc}$ is the number of spikes that coincide within a tolerance $\pm \gamma$, $\langle N_{coinc}\rangle=2R_\sigma\gamma N_{data}$ is the expected number of coincidences for a Poisson spike train with the same rate as the data, and $\mathcal{N}^{-1}=1-2R_{\sigma}\gamma$ is a normalization factor\cite{Kistler1997}. The coincidence factor is zero for random Poisson coincidence and is one for spike trains that agree exactly. 

\subsection*{Information analysis}
When the firing rate is the relevant output, the coding efficiency of a neuron can be quantified by calculating the information transmitted about the input current by the observation of a spike.   As shown in references \cite{Brenner2000,Aguera2003a,Fairhall2006}, the information per spike in bits is
\begin{equation}
I[\sp;i(t)]=\displaystyle\int_0^T\!\frac{dt}{T}\frac{R(t)}{\bar{R}}\log_2\!\left[\frac{R(t)}{\bar{R}}\right], \label{InfoD}
\end{equation}
where $T$ is the duration of the spike train.  For each input standard deviation, Eq. \eqref{InfoD} can be applied directly to the output of the dynamical models when the instantaneous firing rate is sampled with bins of finite duration, giving the information $I_{\sigma}^D[\sp;i(t)]$.  

For the LN models, the information per spike transmitted about the filtered stimulus can be calculated similarly:
\begin{align}
I^{LN}_{\sigma}[\sp;s_x(t)]&=\displaystyle\int\!ds_x\,p_{\sigma}[s_x]\frac{R_{\sigma}[s_x]}{\bar{R}_{\sigma}}\log_2\!\left[\frac{R_{\sigma}[s_x]}{\bar{R}_{\sigma}}\right], \label{InfoLN0}\\
&=\displaystyle\int\!ds_x\,p_{\sigma}[s_x|\sp]\log_2\!\left[\frac{p_{\sigma}[s_x|\sp]}{p_{\sigma}[s_x]}\right], \label{InfoLN}
\end{align}
where the time average has been replaced by an ensemble average over the filtered stimulus; the second equality uses the definition of the rate estimation function from reverse correlation in Eq. \eqref{NonLinFromData}. LN models are reduced descriptions of the dynamics and so, by the data-processing inequality, $I^{LN}\le I^D$. 

The second equality in Eq. \eqref{InfoLN} takes the \emph{decoding} perspective: how much information does the spiking response provide about the stimulus\cite{RiekeSpikes}. 
From the decoding perspective, the ratio of the LN model information to the dynamical model information provides a measure of the completeness of the LN model as a decoding model. When the ratio is close to one, the dynamical model can be thought of as an \emph{implementation} of the code based on linear feature selection. For a decoding showing perfect contrast gain control, the information per spike is independent of $\sigma$, as is easily shown by changing variables to $s_x/\sigma$ in Eq. \eqref{InfoLN}.

\subsection*{Filtered stimulus-conditional input ensembles}
In this section, we summarize the necessary mathematical machinery to
work with arbitrary filtered stimuli in the context of driven dynamical systems.  In Eq. \eqref{s_x}, we define the filtered stimulus as convolution with a filter $h_x(t-t')$, and so filtered stimuli are Gaussian processes \cite{Hida1993}. Throughout this paper, we assume that inputs are statistically in steady state ($t\gg 0$).  To calculate LN models from first principles, we need to know the statistics of the white noise input when conditioned on the observation of a particular value of the filtered input. We will first characterize the conditional mean and autocorrelation functions for the input current, $\langle i(t')|s_x(t)\rangle$ and $\phi_{ii|s_x}(t',t'';t)$.  

Given a filtered observation $s_x(t)$, one can invert (deconvolve) the filtering to reconstruct the original input current, $i(t)$.  In continuous time, the inverse convolution operator, $\hat{h}_x^{-1}(t-t')$, is the differential operator defined by:
\begin{align}
\displaystyle \int_0^t \frac{dt'}{\tau}\,\hat{h}_x^{-1}(t-t') h_x(t')=\tau\delta(t-t') \label{InverseFilterDefinition}
\end{align} 
and so performs the operation:
\begin{align}
i(t)=\displaystyle \int_0^t \frac{dt'}{\tau} \hat{h}_x^{-1}(t-t') s_x(t'). \label{IofS}
\end{align}
For causal filters whose inverse operator can be expressed as a power series in $\frac{d\,}{dt}$, the left and right inverses are equal:
\begin{align}
\displaystyle \int_0^t \frac{dt'}{\tau}\,\hat{h}^{-1}(t-t') h(t')=\displaystyle \int_0^t \frac{dt'}{\tau}\, h(t') \hat{h}^{-1}(t-t') \label{RightInverse}
\end{align} 
 as can be verified by integrating by parts over a test function.  For an arbitrary filter, one can construct the inverse explicitly\cite{Hirschman1950,Hibino1993}, but here it is sufficient to work only with the special case of an exponential filter.  For normalized exponential filters with dimensionless inverse time constant $k$,
\begin{equation*}
h(t-t')=\sqrt{2k}e^{-\frac{k(t-t')}{\tau}}\text{H}(t-t'), 
\end{equation*}
the inverse convolution operator is
\begin{equation}
\hat{h}^{-1}(t-t')=\frac{1}{\sqrt{2k}}\left(k+\tau\frac{d}{dt}\right)\tau\delta(t-t'),\label{InverseFilterExp}
\end{equation}
as can be verified by substitution into Eq. \eqref{InverseFilterDefinition}.  

From Eq. \eqref{IofS}, we can get the properties of $i(t')$ given $s_x(t)$ once we have the conditional mean and autocorrelation of the filtered stimulus.  The mean conditional moment is most easily inferred directly from the well-known autocorrelation function of a Gaussian process\cite{Hida1993}, using the Law of Total Expectation identity, 
$\langle s_x(t')s_x(t)\rangle=\big\langle \langle s_x(t')|s_x(t)\rangle s_x(t)\big\rangle\equiv \phi_{s_xs_x}(t',t'')$, giving:
\begin{equation}
\langle s_x(t')|s_x(t)\rangle=s_x(t)\displaystyle \int_0^{\text{min}(t,t')}\frac{dt_1}{\tau}h_x(t-t_1)h_x(t'-t_1).  \label{MeanSS}
\end{equation}
Similarly, the conditional autocorrelation function, is most easily inferred from the Law of Total Covariance identity, $\phi_{s_xs_x|s_x}(t',t'';t)=\phi_{s_xs_x}(t',t'')-\big\langle \langle s_x(t')|s_x(t)\rangle \langle s_x(t'')|s_x(t)\rangle \big\rangle $, and is:
\begin{align}
\phi_{s_xs_x|s_x}(t',t'';t)&=\sigma^2\left[ \displaystyle \int_0^{\text{min}(t',t'')}\frac{dt_1}{\tau}h_x(t'-t_1)h_x(t''-t_1) \right. \label{PhiSSS}\\
&\qquad \qquad- \left. \int_0^{\text{min}(t,t')}\frac{dt_1}{\tau}h_x(t-t_1)h_x(t'-t_1) \int_0^{\text{min}(t,t'')}\frac{dt_1}{\tau}h_x(t-t_1)h_x(t''-t_1) \right]. \notag
\end{align}
Note that the conditional autocorrelation does not depend on the value of $s_x(t)$. Higher-order correlation functions may be constructed from the first two with the usual conditional Gaussian closure relations\cite{Billingsley1995}. 

Using the moments of the filtered stimulus, one can now compute the properties of the original white noise input ensemble conditioned on observing a particular value of the filtered stimulus.  Using Eqs. \eqref{IofS} and \eqref{MeanSS}, the conditional mean current is
\begin{align}
\langle i(t')|s_x(t)\rangle &= \displaystyle \int_0^{t'}\frac{dt_1}{\tau}\hat{h}^{-1}(t'-t_1) \langle s_x(t_1)|s_x(t)\rangle, \notag \\ 
&= s_x(t)h_x(t-t'). \label{MeanIS}
\end{align}
Since the filter is causal, $\langle i(t')|s_x(t)\rangle$ is only non-zero for $t'<t$.  Similarly, we can find the conditional autocorrelation function with Eq. \eqref{PhiSSS}:
\begin{align}
\phi_{ii|s_x}(t',t'';t)=\sigma^2\left[\tau\delta\!\left(t'-t''\right)-h_x(t-t')h_x(t-t'')\right]. \label{PhiIIS}
\end{align}
Eqs. \eqref{MeanIS} and \eqref{PhiIIS} define an $s_x$-dependent Gaussian source with discrete time conditional distribution:
\begin{equation}
p\big[i(t')\big| s_x(t)\big]=\frac{1}{\sqrt{2\pi\sigma^2\frac{\tau}{dt}\left(1-h_x(t-t')^2\frac{dt}{\tau}\right)}}e^{-\frac{\left(i(t')-s_x(t)h_x(t-t')\right)^2}{2\sigma^2\frac{\tau}{dt}\left(1-h_x(t-t')^2\frac{dt}{\tau}\right)}} \label{PIS}
\end{equation} 
These expressions capture the bias in the mean and the reduction in variance due to selecting out the component proportional to $h_x(t)$ and leaving all other components unobserved.

\subsection*{Dynamics conditioned on the filtered stimulus}
While the dynamics are deterministic because the realization of the input is assumed to be known, in transitioning to the LN model framework, stochasticity is introduced by throwing away the influence of all history not captured by the filter. The stochastic dynamical process that is statistically equivalent to the LN model with a given filter, $h_x$, is the one that describes the evolution of the voltage preceding an observed target value of the filtered stimulus.   This process evolves according to a conditional transition probability, $p\big[v(t')\big| v(t'-dt),s_x(t)\big]$. We develop this section for the EIF model.  LIF-specific results follow in the $\Delta\rightarrow 0$ limit.

We find the conditional transition probability by marginalizing over the input current given $s_x(t)$:
\begin{align}
p\big[v(t')\big| v(t'-dt),s_x(t)\big]=\displaystyle \int di(t'-dt)\,p\big[v(t')\big| v(t'-dt),i(t'-dt)\big]p\big[i(t'-dt)|s_x(t)\big], \notag
\end{align}
where $p\big[i(t'-dt)\big|s_x(t)\big]$ is given by Eq. \eqref{PIS} and $p\big[v(t')\big| v(t'-dt),i(t'-dt)\big]$ is given in Eq. \eqref{LIFPropagator}. 
The conditional transition probability of the process is
\begin{align}
p\big[v(t')\big|& v(t'-dt),s_x(t)\big]=\notag \\
&\frac{\text{H}\big[v_{s}-v(t'-dt)\big]}{\sqrt{2\pi\sigma^2\frac{dt}{\tau}\left(1-h_x(t-t'+dt)^2\frac{dt}{\tau}\right)}}e^{-\frac{\left(v(t')-v(t'-dt)-\frac{dt}{\tau}\!\left[-v(t'-dt)+v_o+f\left(v(t'-dt)\right)+s_x(t)h_x(t-t'+dt)\right]\right)^2}{2\sigma^2 \frac{dt}{\tau}\left(1-h_x(t-t'+dt)^2\frac{dt}{\tau}\right)} } \notag \\
&\quad +\quad\frac{\text{H}\big[v(t'-dt) -v_{s}\big]}{\sqrt{2\pi\sigma^2\frac{dt}{\tau}\left(1-h_x(t-t'+dt)^2\frac{dt}{\tau}\right)}}e^{-\frac{\left(v(t')-v_r-\frac{dt}{\tau}\!\left[-v_r+v_o+f\left(v_r\right)+s_x(t)h_x(t-t'+dt)\right]\right)^2}{2\sigma^2 \frac{dt}{\tau}\left(1-h_x(t-t'+dt)^2\frac{dt}{\tau}\right)} } . \label{ConditionalProcess}
\end{align}
Eq. \eqref{ConditionalProcess} in principle gives complete information about the evolution of the ensemble of voltage trajectories given an observed value of the filtered stimulus. 

To compute the rate estimation function in Eq. \eqref{NonLinFromVoltageEIF0}, we need $p\big[v(t),v(t-dt)\big|s_x(t)\big]$.  We factor this into two parts:
\begin{align*}
p\big[v(t),v(t-dt)\big|s_x(t)\big]=p\big[v(t-dt)\big|v(t),s_x(t)\big]p\big[v(t)\big|s_x(t)\big].
\end{align*}
The first term is the backward conditional transition probability.  Because of the reset, there are two ways $v(t)$ can be arrived at from $v(t-dt)$: directly from nearby voltages or indirectly via the instantaneous reset.  The backward conditional transition probability is thus of the form
\begin{equation}
p\!\big[v(t'-dt)\big| v(t'),s_x(t)\big]=\frac{ p^{(0)}\!\big[v(t'-dt)\big| v(t'),s_x(t)\big]+p^{(0)}\!\big[v(t'-dt)\big| v(t')\ge v_s,s_x(t)\big]p^{(0)}\!\big[v_r\big| v(t'),s_x(t)\big]}{\mathcal{N}\!\left[v(t')\right]}, \label{BackwardConditionalProcess}
\end{equation}
where $\mathcal{N}\!\left[v(t')\right]$ is the normalization constant such that $\int\! dv(t'-dt)\,p\big[v(t'-dt)\big| v(t'),s_x(t)\big]=1$, and $p^{(0)}\!\big[v(t'-dt)\big| v(t'),s_x(t)\big]$ is the free backward conditional transition probability ignoring the reset. 
The forward evolution of the free system during a single time step is statistically reversible because it is Gaussian\cite{Weiss1975}, so the free backward conditional transition probability is the time-reversal of the forward \cite{Kautz1988}:
\begin{align}
p^{(0)}\big[v(t'-dt)\big| v(t'),s_x(t)\big]&\propto e^{-\frac{\left(v(t'-dt)-v(t')-\frac{dt}{\tau}\!\left[-v(t')+v_o+f\left(v(t')\right)+s_x(t)h_x(t-t'+dt)\right]\right)^2}{2\sigma^2 \frac{dt}{\tau}\left(1-h_x(t-t'+dt)^2\frac{dt}{\tau}\right)} }. \label{BackwardConditionalProcess0}
\end{align}
In the domain of integration required for the rate estimation function, $v(t)\ge v_{th,\sigma}$, the second term in Eq. \eqref{BackwardConditionalProcess} is always negligible when $v_{th,\sigma}-v_r\gg \sigma\sqrt{dt/\tau}$.

The voltage estimation distribution, $p\big[v(t)\big| s_x(t)\big]$, can formally be found by repeated application of the the forward conditional transition probability evolved from the steady-state voltage distribution in the distant past:
\begin{align}
p\big[v(t)\big| s_x(t)\big]=\displaystyle \int dv(t-dt) \ldots \int dv(0)\,p\big[v(t)\big| v(t-dt),s_x(t)\big]\ldots p\big[v(dt)\big| v(0),s_x(t)\big]p\big[v(0)\big], \label{FormalPVS}
\end{align}
where $p\!\left[v(0)\right]$ is the steady-state distribution in Eq. \eqref{PVSteadyState}. With Eqs. \eqref{NonLinFromVoltageEIF0} and \eqref{RofVEIF}, we have a formal solution in terms of integrals for the rate estimation function given a choice of the filter.  In later sections, we demonstrate some asymptotic results based on moments of $p\big[v(t)\big|s_x(t)\big]$.  

To simplify the expression for the estimated firing rate in Eq. \eqref{NonLinFromVoltageEIF0}, we perform the integral over $v(t-dt)$ using Eq. \eqref{BackwardConditionalProcess} to obtain
\begin{align}
R_{\sigma}\!\left[s_x(t)\right]=\!\frac{1}{dt}\displaystyle \int_{v_{th,\sigma}}^{\infty}\!\!\!\!\! dv(t)\,\frac{1}{2}\text{erfc}\!\left(\frac{v(t)-v_{th,\sigma}+\frac{dt}{\tau}\!\left[-v(t)+v_o+f\left(v(t)\right)+s_x(t)h_x(dt)\right]}{\sqrt{2\sigma^2\frac{dt}{\tau}\left(1-h_x(dt)^2\frac{dt}{\tau}\right)}}\right)p\big[v(t)\big|s_x(t)\big], \notag
\end{align}  
where we assume that $v_{th,\sigma}\ll v_s$. Because the variance in the complementary error function goes to zero with $dt$, the error function only has very narrow support near $v_{th,\sigma}$ in the domain of integration, and so we can replace the error function with a delta function at $v_{th,\sigma}$ under the integral:
\begin{align}
R_{\sigma}\!\left[s_x(t)\right]\approx\frac{A}{dt}\displaystyle \int dv(t)\,\delta\!\left[v(t)-v_{th,\sigma}\right]p\big[v(t)\big|s_x(t)\big]. \notag
\end{align}
The amplitude of the delta-function, $A$, is determined by the integral of the complementary error function above threshold.  Using the definition of the stochastic dynamical threshold in Eq. \eqref{StochasticThreshold2} and keeping only the leading order in $\sqrt{dt/\tau}$, 
\begin{align}
A&=\displaystyle \int_{v_{th,\sigma}}^{\infty} dv(t)\,\frac{1}{2}\text{erfc}\!\left(\frac{v(t)-v_{th,\sigma}+\frac{dt}{\tau}\!\left[-v(t)+v_o+f\left(v(t)\right)+s_x(t)h_x(dt)\right]}{\sqrt{2\sigma^2\frac{dt}{\tau}\left(1-h_x(dt)^2\frac{dt}{\tau}\right)}}\right) \notag \\
 &\approx \displaystyle \int_{v_{th,\sigma}}^{\infty} dv(t)\,\frac{1}{2}\text{erfc}\!\left(\frac{v(t)-v_{th,\sigma}}{\sqrt{2\sigma^2\frac{dt}{\tau}}}+\text{erf}^{-1}\!\left(2C-1\right)\right) \notag \\
&=\beta(C) \sqrt{\frac{\sigma^2 dt}{2\pi\tau}}, \notag
\end{align}
where:
\begin{equation}
\beta(C)=e^{-\!\left(\text{erf}^{-1}\!\left(2C-1\right)\right)^2}+2(C-1)\sqrt{\pi}\text{erf}^{-1}\!\left(2C-1\right). \label{Beta}
\end{equation}
$\beta\!\left(C\right)$ is $\mathcal{O}(1-C)$ for $0.05\lesssim C \lesssim 0.95$.  The final result for rate estimation function for the EIF model is given in Eq. \eqref{NonLinFromVoltageEIF}. As discussed near Eq. \eqref{NonLinFromVoltageEIF}, the coefficient $A$ can be interpreted as defining the effective size of the set of voltages that correspond to a spike time.

The LIF model result presented in Eq. \eqref{NonLinFromVoltageLIF} also follows from this more general development from the EIF model when $\Delta\rightarrow 0$.  In the limit, the stochastic dynamical threshold, $v_{th,\sigma}$ goes to the deterministic threshold, $v_{th}$, for all $\sigma$.  Similarly, the defining equation of the LN model of the EIF model in Eq. \eqref{NonLinFromVoltageEIF0} reduces to the equation for the LIF model in Eq. \eqref{NonLinFromVoltageLIF0} because the integral over $dv(t-dt)$ is always one in the LIF limit.

\subsection*{Adaptation of the optimal filter}
To see how the filtered stimulus acts as a linear estimator of the voltage, we use the ``inverse'' of the trick used in Eq. \eqref{SofD} to put the filtered stimulus into the dynamical system. We formally integrate the dynamical model, as shown in Eq. \eqref{LIFIntegral}, and add and subtract from the right side $\alpha\big(\hat{h}_m^{-1}\ast s_x\big)(t)-\alpha\big(\hat{h}_m^{-1}\ast s_x\big)(t)$, where $\hat{h}_m^{-1}$ is the inverse convolution operator for the membrane filter, from Eq. \eqref{InverseFilterExp}, and $\alpha$ is a normalization constant. Then,
\begin{align}
v(t)&=v_o+\alpha s_x(t)+e(t), \label{LIFEstimation} \\
e(t)&=\displaystyle \int_0^t \frac{dt'}{\tau} e^{\frac{t'-t}{\tau}}\left[i(t')+f\!\left(v(t')\right)-\left(v_{s}-v_r\right)\tau R(t')-\alpha\big(\hat{h}_m^{-1}\ast s_x\big)(t')\right]. \label{ErrorTerm}
\end{align}
Eqs. \eqref{LIFEstimation} and \eqref{ErrorTerm} are exact, but are in the suggestive form of a linear instantaneous estimate of the voltage, $v_o+\alpha s_x(t)$, plus an ``error term,'' $e(t)$, that depends on all of the history of the stimulus and the nonlinear currents up to time $t$.  The dimensionality reduction and corresponding loss of determinism going from the nonlinear spiking system to the LN model is the result of throwing away all history that cannot be accounted for by the instantaneous value of $s_x(t)$.  The optimally predictive filter will ``explain'' as much of the error term as possible when the voltage is at threshold, as represented by the uncertainty reduction due to maximizing the information in Eq. \eqref{InfoLN0}. 

While we cannot analytically solve the information optimization problem in Eq. \eqref{HOpt} outside of the limits presented in the body text, we can gain insight by examining the filter that approximately minimizes the variance of $e(t)$, given that the voltage is below threshold. By definition, minimizing the variance of the error term is equivalent to maximizing the correlation of the voltage and the filtered stimulus since
\begin{align}
\text{Var}\!\left[e(t)\big|{v(t)\le v_{th,\sigma}}\right]&=\text{Var}\!\left[v(t)-\alpha s_x(t)\big|{v(t)\le v_{th,\sigma}}\right], \notag\\
&=\text{Var}\!\left[v(t)\big|{v(t)\le v_{th,\sigma}}\right]+\text{Var}\!\left[\alpha s_x(t)\big|{v(t)\le v_{th,\sigma}}\right]-\text{Cov}\!\left[v(t)\alpha s_x(t)\big|{v(t)\le v_{th,\sigma}}\right], \notag
\end{align}
and $\alpha$ is chosen so that $\text{Var}\!\left[\alpha s_x(t)\big|{v(t)\le v_{th,\sigma}}\right]=\text{Var}\!\left[v(t)\big|{v(t)\le v_{th,\sigma}}\right]$.
We find the filter that minimizes the variance of the error term  by using a self-consistent approximation called \emph{stochastic linearization} that we have treated previously\cite{Famulare2010}.  We denote the stochastic linearization filter $h_l(t)$, and it solves the optimization problem:
\begin{equation}
h_l(t)=\operatorname*{arg\,max}_{h_x(t)} \, \text{Var}\!\left[e(t)\big|{v(t)\le v_{th,\sigma}}\right],
\end{equation}
under some assumptions.

\subsubsection*{Stochastic linearization filter for the LIF model}
We start with the LIF model because all of its dynamical evolution takes place below threshold and so  $\text{Var}\!\left[e(t)\big|v(t)\le v_{th,\sigma}\right] =  \text{Var}\!\left[e(t)\right]$. 
For one-dimensional dynamical models, the stochastic linearization approximation looks for the most predictive exponential filter with a $\sigma$-dependent time constant.  We expect there is a good single exponential approximation to the optimal filter for the following reasons. The dynamical models are stochastic processes, and all correlation functions can in principle be derived from the corresponding Fokker-Planck equation.  As the Fokker-Planck equation is linear, it can be decomposed into eigenmodes that are a function of the input standard deviation, and all statistical properties follow from the eigenmodes\cite{Risken}.  For example, steady-state statistics follow from the eigenmode with zero eigenvalue.  For excitable one-dimensional models, the eigenvalues are real and the spectrum is discrete\cite{Knight2000}.  Thus, low-order statistics have long time behavior dominated by the lowest non-zero eigenvalue, and the optimal filter must account for those statistics. So, on general grounds, we expect the optimal filter to have exponential long-time behavior. 

We search for an exponential filter with dimensionless inverse time constant $k_{\sigma}$.  
For an exponential filter, we can make a substitution using an identity that follows immediately from Eqs. \eqref{IofS} and \eqref{InverseFilterExp}: $i(t)=\tau\alpha \dot{s}_{v}+ k_{\sigma} \alpha s_v$, where $\alpha$ is a normalization constant. The variance of the error term, Eq. \eqref{ErrorTerm}, is:
\begin{align}
\text{Var}[e(t)]&=\displaystyle \int^t \!\frac{dt'}{\tau}\int^t \!\frac{dt''}{\tau} e^{\frac{t'+t''-2t}{\tau}}\bigg[ (k_{\sigma}-1)^2\alpha^2\left\langle s_v(t')s_v(t'')\right\rangle   +\left(v_{th}-v_{r}\right)^2\tau^2\phi_{RR}(t'-t'')\ldots \notag\\
& \qquad \qquad  \qquad  \qquad  \qquad  \qquad -2 (k_{\sigma}-1) \left(v_{th}-v_{r}\right)\tau\alpha\left\langle R(t')s_v(t'')\right\rangle\bigg], \label{VarErrorLIF}
\end{align}
where $\phi_{RR}(t'-t'')=\left(\left\langle R(t')R(t'')\right\rangle -\bar{R}^2\right)$ is the rate autocorrelation function.  Unfortunately, the correlation functions involving the rate are unknown.

In stochastic linearization, this closure problem is solved by assuming that the variance of the error term is identically zero. The approximation proceeds in two steps. First, since $\text{Var}[e(t)]=\text{Var}[v(t)-\alpha s_v(t)]$, under the assumption that the variance of the error term is identically zero, it follows that $\alpha s_v(t)=v(t)-\langle v \rangle$.  Second, one assumes that all the correlation functions that appear in Eq. \eqref{VarErrorLIF} have the same time dependence, which we denote $g(t'-t'')$; this assumption is good when the correlation functions are dominated by the lowest non-zero eigenmode discussed previously.
Under these assumptions,
\begin{align*}
\alpha^2\left\langle s_v(t')s_v(t'')\right\rangle&=\left(\langle v^2\rangle - \langle v \rangle^2\right) g(t'-t''), \\
\alpha\left\langle R(t')s_v(t'')\right\rangle&=\left(\langle R v\rangle - \bar{R}_{\sigma}\langle v \rangle\right) g(t'-t'')\\
\phi_{RR}(t'-t'')&=\left(\langle R^2\rangle - \bar{R}_{\sigma}^2\right) g(t'-t'').
\end{align*}
The rate-voltage cross correlation is $\langle R v\rangle=\bar{R}_{\sigma}\langle v|\sp\rangle = \bar{R}_{\sigma}v_{th}$.  In this approximation, the variance of the error term is
\begin{align}
\text{Var}[e(t)]\propto& \bigg[ (k_{\sigma}-1)^2\left(\langle v^2\rangle - \langle v \rangle^2\right) -2 (k_{\sigma}-1) \left(v_{th}-v_{r}\right)\tau\bar{R}_{\sigma}\left(v_{th}-\langle v\rangle\right) \ldots   \notag \\
&\qquad \qquad \qquad +(v_{th}-v_r)^2\tau^2\left(\langle R^2\rangle - \bar{R}_{\sigma}^2\right)\bigg] \displaystyle\! \int^t \!\frac{dt'}{\tau}\int^t \!\frac{dt''}{\tau} e^{\frac{t'+t''-2t}{\tau}} g(t'-t''). \label{VarErrorLIF2}
\end{align}
Minimization with respect to $k_{\sigma}$ gives the result in Eq. \eqref{KslLIF}. 

For one-dimensional models, the stochastic linearization filter differs from the exponential membrane filter.     For the LIF model, it is important to realize that the membrane filter, corresponding to $k=1$, is not in the eigenspectrum for finite $\sigma$ because of the reset nonlinearity (or equivalently, because of the non-natural boundary conditions in the Fokker-Planck formulation) and thus does not appear in the relevant correlation functions.  This explains why the membrane filter is sub-optimal for predicting the response for larger input standard deviations.  

\begin{figure}[!ht]
\begin{center}
\includegraphics{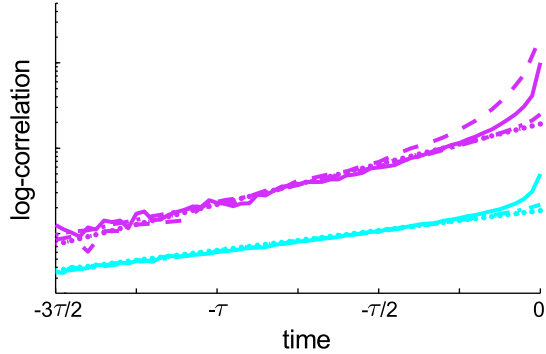}
\end{center}
\caption{ {\bf Stochastic linearization predicts the lowest non-zero eigenmode of the driven dynamics.}  We show second-order correlation functions for the LIF model, on a log-linear plot with the y-axis in arbitrary units, scaled and shifted for comparison. Rate autocorrelation function (dashed), spike-triggered average voltage (dash-dot), STA current (solid), and the stochastic linearization filter (dotted).  The upper data set corresponds to $\sigma=8$, with $k_{8}=2.5$ and the lower, $\sigma=0.45$ with $k_{0.45}=1.06$. The long time behavior the correlation functions is governed by the timescale identified in Eq. \eqref{KslLIF}.  The decay of the spike-triggered voltage reflects the decorrelation of the voltage with the firing rate over longer timescales. }
\label{LowEigenModeFig}
\end{figure}

As anticipated by the eigenmode argument, the long time behavior of the low-order correlation functions is given by the timescale defined in Eq. \eqref{KslLIF} as can be seen from the spike-triggered average current, the spike-triggered voltage, and the rate autocorrelation function. This is verified in Fig. \ref{LowEigenModeFig}.

The reader surely noticed some sleight of hand in the minimization of Eq. \eqref{VarErrorLIF2}.  The filtered stimulus $s_v$ is an OU-process, and so $g(t'-t'')=e^{-\frac{k_{\sigma}|t'-t''|}{\tau}}$.  Why do we not minimize over the $k_{\sigma}$ dependence of the integrated $g(t'-t'')$?  It is incorrect to vary over the $k_{\sigma}$-dependence in $g$ because doing so introduces a purely multiplicative degree of freedom into the minimization of $\text{Var}[e(t)]$. This is equivalent to assuming that the variance of the voltage is undetermined prior to minimization. In lieu of introducing a Lagrange multiplier to constrain the multiplicative scaling, we held the degree of freedom fixed.  Discussion of this issue and related subtleties has been extensive in the stochastic linearization literature. References \cite{Crandall2001,Socha2005p1,Crandall2006} are among the more accessible; our approach is `true' linearization as described therein. 

\subsubsection*{Semi-empirical closed form for the STA of the LIF model}
At short times prior to the spike for small time steps $dt$, Paninski has shown\cite{Paninski2006IFSTA} (see also Badel \emph{et al.}\cite{Badel2006}) that the STA has a square-root singularity at short times that, in our notation, is
\begin{equation*}
\langle i(t)|\sp\rangle= \sigma\sqrt{\frac{2}{\pi}}\left(\frac{dt}{-t}\right)^{1/2} \qquad \text{for }  -\tau\ll -t<0.
\end{equation*}
This component of the STA is entirely determined by the threshold boundary and is independent of the subthreshold dynamics. 
For long times prior to the spike, the stochastic linearization technique describes the noise-modified effective linear dynamics, an OU-process: $\tau \dot{v}=-k_{\sigma}(v-\langle v\rangle)+I(t)$.    If there was no boundary effect, the STA of the effective linear dynamics (the average input current conditioned on hitting $v_{th}$) would be:
\begin{equation*}
\langle i(t)|\sp\rangle=2\left(v_{th}-\langle v\rangle\right)e^{\frac{k_{\sigma}t}{\tau}} \qquad \text{for } t\ll -\tau,
\end{equation*}
as follows immediately from Eq. \eqref{IofS} with appropriate normalization. 

 The method of matched asymptotic solutions\cite{Verhulst2005} can provide an accurate closed form approximation to the STA for all times.  In this method, each solution is used where it is valid, and a complete solution is found by matching the two at an empirically chosen intermediate point, $t_m$. Since the boundary term behavior is power-law, it decays very slowly relative to the exponential long-time behavior and so it needs to be cut off at the matching point.  The amplitude of the exponential component is determined by the amount of linear subthreshold integration required to bring the voltage near enough to threshold so that a fast fluctuation in few time steps can bring the voltage above threshold.  With the aid of simulation, we find that an excellent fit to the STA for sufficiently small $dt$ is given by:
\begin{align}
\left\langle i(t)\big|\sp\right\rangle= \left\{
\begin{array}{ll}
a_{\sigma}e^{\frac{k_{\sigma}(t)}{\tau}} & \text{for } t<t_m,\\
 a_{\sigma}e^{\frac{k_{\sigma}(t)}{\tau}} +\sigma\sqrt{\frac{2}{\pi}}\!\left[\left(\frac{dt}{-t}\right)^{\frac{1}{2}}-\left(\frac{dt}{t_m}\right)^{\frac{1}{2}}\right] & \text{for } t_m\le t<0,\\
0 & \text{for } t\ge 0.
\end{array}
\right. \label{StaAn}
\end{align}
Coefficient $a_{\sigma}$ and the optimal matching time depend on the ratio of  $v_{th}-\langle v\rangle$ to $\sigma \sqrt{\tau/dt}$.  In typical simulations,  $\sigma\sqrt{\frac{\tau}{dt}}\gg v_{th}-\langle v\rangle$ always.

\begin{figure}[!ht]
\begin{center}
\includegraphics{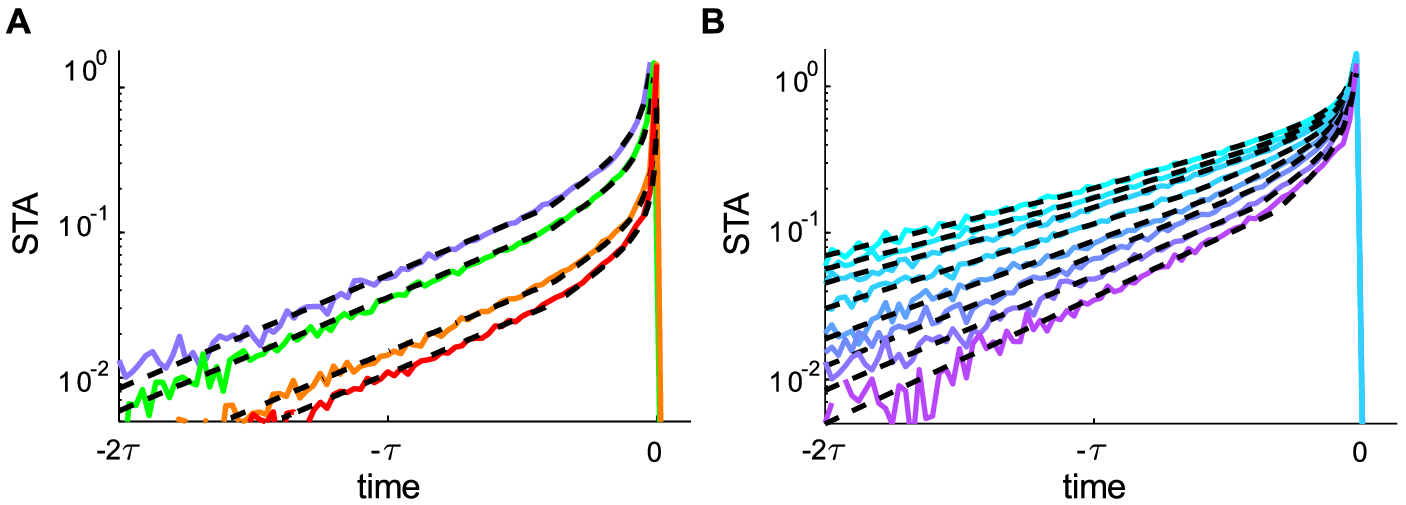}
\end{center}
\caption{ {\bf Semi-empirical closed form for the STA of the LIF model}. The result in Eq. \eqref{StaAnDt} is shown (dashed black) over the corresponding simulation result for typical examples; y-axis is STA in terms of $\frac{\langle i(t)|\sp\rangle}{\sigma}$. \textbf{A}. Fixed standard deviation, varying time step: $\sigma=2$, $\frac{dt}{\tau}=\{40^{-1},80^{-1},400^{-1},800^{-1}\}$: purple, green, orange, red respectively. $G[x]=1.9\frac{\left(v_{th}-\langle v\rangle\right)}{\sigma\sqrt{\frac{\tau}{dt}}}$, $\frac{t_m}{\tau}=\frac{3}{8}$. \textbf{B}. Fixed time step $\frac{dt}{\tau}=40^{-1}$, varying standard deviation.  Color code as in previous figures.  }
\label{LIFStaAnFig}
\end{figure}

For larger $dt$ (although still $dt\ll \tau$), Eq. \eqref{StaAn} fails even though the qualitative behavior of the STA remains the same: empirically, the square-root singularity is modified. An excellent fit can still be found if we allow the exponent and matching time to depend on the time step.  We find that the general semi-empirical closed form is
\begin{align}
\left\langle i(t)\big|\sp\right\rangle= \left\{
\begin{array}{ll}
a_{\sigma}e^{\frac{k_{\sigma}(t)}{\tau}} & \text{for } t<t_m,\\
a_{\sigma}e^{\frac{k_{\sigma}(t)}{\tau}} +\frac{\sigma}{\sqrt{\pi}}\!\left[\left(\frac{2dt}{-t}\right)^{\nu}-\left(\frac{2dt}{t_m}\right)^{\nu}\right] & \text{for } t_m\le t<0,\\
0 & \text{for } t\ge 0.
\end{array}
\right. \label{StaAnDt}
\end{align}
Coefficient $a_{\sigma}$ scales as
\begin{align}
a_{\sigma}\rightarrow\left\{ 
\begin{array}{ll}
2\left(v_{th}-v_o\right) & \text{for } \sigma\rightarrow 0,\\
\sigma\, G\!\left[\frac{\left(v_{th}-\langle v\rangle\right)}{\sigma\sqrt{\frac{\tau}{dt}}}\right] & \text{for } \sigma\sqrt{\frac{\tau}{dt}}\gg v_{th}-\langle v\rangle,
\end{array}
\right. \notag %\label{ASigma}
\end{align} 
with $G[x]\sim 2\frac{\left(v_{th}-\langle v\rangle\right)}{\sigma\sqrt{\frac{\tau}{dt}}} $ and $\frac{t_m}{\tau}\sim 2.2\left(G[x]-0.1\right)$ for $\frac{\left(v_{th}-\langle v\rangle\right)}{\sigma\sqrt{\frac{\tau}{dt}}}<0.4$ empirically; the numerical coefficient in $G[x]$ is weakly $\sigma$-dependent.  
The exponent scales logarithmically as
\begin{equation}
\nu\approx \frac{1}{2}+0.15\,\text{ln}\!\left(1+660\frac{dt}{\tau}\right), \notag %\label{Nu}
\end{equation}
although we find that the fit is indistinguishable for $\nu\pm 0.1$.\footnote{For $dt=\tau/800$, the observed logarithmic scaling predicts $\nu\approx 0.6$. In this case, change in the goodness of fit for the small correction from $\nu=1/2$ can easily be compensated by small changes in the amplitude and matching time.}   For $\sigma\sqrt{\frac{dt}{\tau}}$ too large, the STA begins to oscillate below this solution for small times (not shown).  
While the result in Eq. \eqref{StaAn} in the limit $dt\rightarrow 0$ stands on elementary theoretical grounds, we are not aware of a principled argument that produces the anomalous scaling of the exponent with $dt$.  We share this result to add it to the cabinet of curiosities associated with the LIF model. Comparison with typical simulation results is shown in Fig. \ref{LIFStaAnFig}.

\subsubsection*{Stochastic linearization filter for the EIF model}
For the EIF model, the derivation of the stochastic linearization filter is more subtle because of the condition that we only optimize over $v(t)\le v_{th,\sigma}$. 
To simplify the error term in Eq. \eqref{ErrorTerm}, the spike-generating current, $f(v)$, is broken into two parts.  Below the dynamical threshold, $v_{th}$, the detailed shape of $f(v)$ is critical for spike initiation and the integration of the input.  However, above $v_{th}$, $f(v)$ drives the spike rapidly and then the system is returned below threshold by the reset; super-threshold, the details of $f(v)$ are largely irrelevant to the evolution of the voltage between spikes.  Previous spikes affect the instantaneous state of the voltage through low pass filtering with the membrane, and so the net effect of the super-threshold currents is to displace the voltage on average by an amount proportional to the distance between rest and threshold and proportional to the mean firing rate: 
\begin{align*}
\displaystyle \int_0^t \frac{dt'}{\tau} e^{\frac{t'-t}{\tau}}\left[f\!\left(v(t')\right)\text{H}\!\left[v(t')-v_{th}\right]-\left(v_{s}-v_r\right)\tau R(t')\right]\approx - \left(v_{th}-v_r\right)\tau\bar{R}_{\sigma},
\end{align*}
and so, inside the above integral, we can take:
\begin{align*}
f\!\left(v(t')\right)\text{H}\!\left[v(t')-v_{th}\right]-\left(v_{s}-v_r\right)\tau R(t')\approx - \left(v_{th}-v_r\right)\tau R(t').
\end{align*}
For voltages below $v_{th}$, the form of $f(v)$ is relevant and so we cannot simplify it. Thus, the subthreshold error term is approximately:
\begin{equation}
e(t)\bigg|_{v(t)\le v_{th,\sigma}}\approx \displaystyle \int_0^t \frac{dt'}{\tau} e^{\frac{t'-t}{\tau}}\left[i(t')+f\!\left(v(t')\right)\text{H}\!\left[v_{th}-v(t')\right]-\left(v_{th}-v_r\right)\tau R(t')-\alpha\big(\hat{h}_m^{-1}\ast s_x\big)(t')\right]. \label{ErrorTermEIF}
\end{equation}
The stochastic linearization filter can now be found from minimizing the variance of Eq. \eqref{ErrorTermEIF} as for the LIF model.  The result is shown in Eq. \eqref{KslEIF0} and an example is shown in Fig. \ref{LNEIFFig}B.

\subsection*{Moment-based asymptotic results}
Because of the non-Markov structure of the conditional voltage process, the formal solution for $p\!\big[v(t)\big|s_x(t)\big]$ in Eq. \eqref{FormalPVS} is difficult to use.  Useful asymptotic results are more accessible from studying the moments of $p\!\big[v(t)\big|s_x(t)\big]$ directly.  Using Eq. \eqref{LIFIntegral} and the conditional input ensemble, we can study the lower moments; the conditional mean,
\begin{align}
\left\langle v(t)\big|s_x(t)\right \rangle &= v_o+\displaystyle \int_{0}^{t} \!\frac{dt'}{\tau} e^{\frac{t'-t}{\tau}}\left[\big\langle i(t')\big| s_x(t)\big\rangle+ \big\langle f\!\left(v(t')\right)\big| s_x(t)\big\rangle -\left(v_{s}-v_{r}\right)\tau\big\langle R(t')\big|s_x(t)\big\rangle\right], \label{MeanVS}
\end{align}
and the conditional variance,
\begin{align}
\text{Var}\left[v(t)\big|s_x(t)\right]&=\displaystyle \int_{0}^{t} \frac{dt'}{\tau}\int_{0}^{t} \frac{dt''}{\tau} e^{\frac{t'+t''-2t}{\tau}}\bigg[ \phi_{ii|s_x}(t',t'';t) + \phi_{ff|s_x}(t',t'';t)+\left(v_{s}-v_{r}\right)^2\tau^2\phi_{RR|s_x}(t',t'';t) \notag \\
&\qquad \qquad+ \phi_{if|s_x}(t',t'';t) - 2\left(v_{s}-v_{r}\right)\tau \phi_{fR|s_x}(t',t'';t) - 2\left(v_{s}-v_{r}\right)\tau \phi_{Ri|s_x}(t',t'';t)\bigg], \label{VarVS}
\end{align}
where $\phi_{ii|s_x}(t',t'';t)$ is given in Eq. \eqref{PhiIIS}. All correlation functions are defined with mean subtraction as in Eq. \eqref{PhiSSS}: $\phi_{RR|s_x}(t',t'';t)$ and $\phi_{ff|s_x}(t',t'';t)$ are stimulus conditioned rate and $f(v)$ autocorrelation functions;  $\phi_{if|s_x}(t',t'';t)$, $\phi_{fR|s_x}(t',t'';t)$, and $\phi_{Ri|s_x}(t',t'';t)$ are the stimulus conditioned cross-correlation functions.  Higher moments in principle follow similarly.  In general, the moments are no more useful than formal solution for the conditional process in Eq. \eqref{ConditionalProcess} because of the unknown correlation functions. However, in limiting cases, analytic results can be derived from the low moments alone.

\subsubsection*{LIF model: small input standard deviation}
In the low variance limit, spikes are well-separated, and in the run up to a spike, previous spikes are rare for most filtered stimuli.  In this limit, $p\!\left[v(t)\big|s_x(t)\right]$ is approximately Gaussian for non-spiking $s_x(t)$, where almost all the variance in the voltage estimate is due to the difference between the optimally predictive filter and the membrane filter of the LIF model.  In this limit, the leading order conditional mean and variance in Eq. \eqref{MeanVS} and \eqref{VarVS} are (suppressing $t$):
\begin{align}
\left\langle v\big|s_x\right\rangle_{\sigma\rightarrow 0}&\approx v_o+  s_x(t)\frac{\left(h_m\ast h_x\right)}{\sqrt{2}}, \notag  \\%\label{MeanVSLowSig} 
\text{Var}_{\sigma\rightarrow 0}\left[v\big|s_x\right]&\approx \frac{\sigma^2}{2}\left[1- \left(h_m\ast h_x \right)^2\right], \notag%\label{VarVSLowSig}  
\end{align}
where $h_m$ is the intrinsic membrane filter defined in Eq. \eqref{MembraneFilter}.  For $\sigma\rightarrow 0$, the variance is minimized when $h_x=h_m$, reproducing the result in Eq. \eqref{HOptLowSigLIF}.  

In the small noise limit, for values of the filtered stimulus $s_m\le s_{th}$, a Gaussian (limiting to a delta-function) based on the above moments describes $p\!\left[v\big|s_x\right]$. For small-but-finite $\sigma$, the Gaussian model will be perturbed because of spiking, taking the super-threshold probability predicted by the Gaussian model and spreading it near the reset, $v_{r}$.  Thus, for small $\sigma$, the full conditional distribution will look like:
\begin{align}
p_{\sigma\rightarrow 0}\!\big[v\big|s_x\big] &\approx
\frac{1}{\sqrt{2\pi\text{Var}_{\sigma\rightarrow 0}\left[v|s_x\right]}}e^{-\frac{\left(v- \left\langle v|s_x\right\rangle_{\sigma\rightarrow 0}\right)^2}{2\text{Var}_{\sigma\rightarrow 0}\left[v|s_x\right]}}\text{H}\left[v_{th}-v\right]+\text{residual}\!\left[v\big| s_x\right], \label{PVSLowSig}
\end{align}
where the residual accounts for the probability that has been reset and depends on the ignored rate-current and rate-rate correlation functions.  Empirically, this formula holds through the onset of spiking at $s_{th}$ with low probability residual. For increasingly large values of the filtered stimulus, more than one spike is possible in the correlation time of the filter and so this relation, based on ignoring the spike autocorrelations, breaks down.  

\begin{figure}[!ht]
\begin{center}
\includegraphics{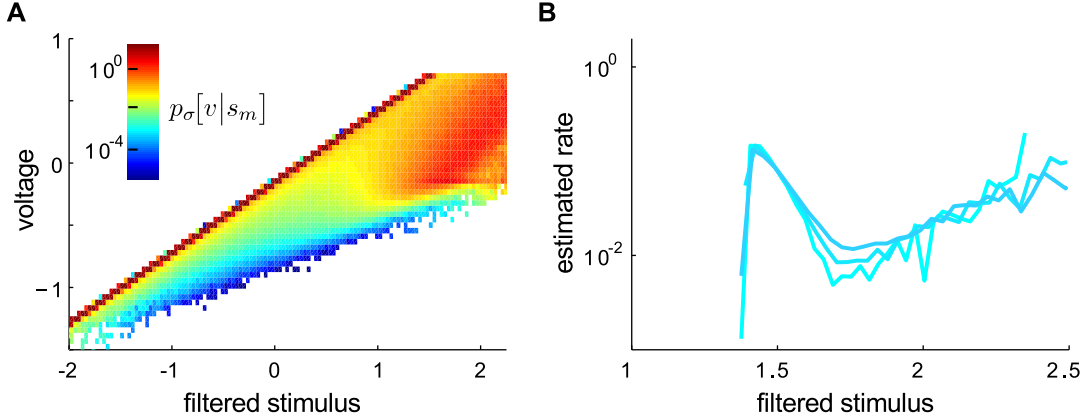}
\end{center}
\caption{ {\bf Asymptotic results small $\sigma$: threshold detection in the LIF model}  \textbf{A}. Numerical conditional voltage distribution for $\sigma=0.45$ for the membrane filter, $p_{\sigma}\!\left[v\big|s_m\right]$, axes in units of $v_{th}-v_o$.  Color shows shows probability density.  In the regime below the threshold filtered stimulus, the residual is two or more orders of magnitude below the primary (singular) Gaussian component. \textbf{B}. Rate estimation functions for the membrane filter, $R_{\sigma}\!\left[s_m\right]$, for $\sigma=\{0.45,0.5,0.55\}$.  In this regime, the LIF model is primarily a detector of the passage to threshold with $s_{th}=\sqrt{2}\left(v_{th}-v_o\right)$. For larger stimuli, the origin of the  fall-off and return to spiking with increasing $s_m$ can be seen in the voltage distribution in panel A as the mass of the prediction at large $s_m$ cycles between $v_{r}$ and $v_{th}$. }
\label{NonLinLowSigFig}
\end{figure}

The rate estimation function following from Eq. \eqref{PVSLowSig} follows from the general solution in Eq. \eqref{NonLinFromVoltageEIF}.  As discussed near Eq. \eqref{HOptLowSigLIF}, the rate estimation functions for the membrane filter are sharply peaked at $s_{th}$ and are approximately independent of the input statistics for $\sigma\lesssim 0.6$.  Results are shown in Fig. \ref{NonLinLowSigFig}.

\subsubsection*{LIF model: large input standard deviation}
\begin{figure}[!ht]
\begin{center}
\includegraphics{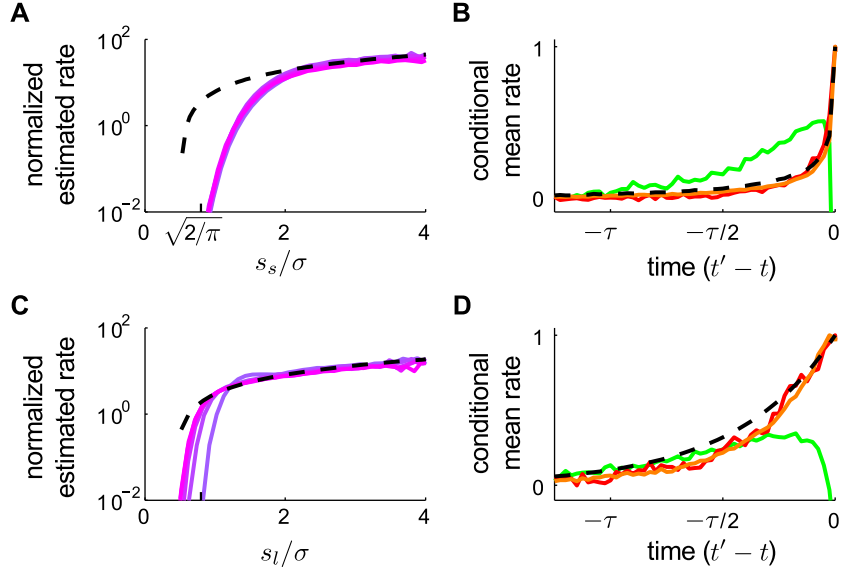}
\end{center}
\caption{ {\bf Asymptotic results at large $\sigma$: perfect contrast gain control in the LIF model}.  \textbf{A}. Normalized rate estimation functions based on the STA filter, and \textbf{C}, stochastic linearization filter, $h_l$, for $4\le \sigma\le 10$.  Asymptotic prediction from Eq. \eqref{NonLinHighSig} is dashed.  The limiting perfect contrast gain control property of the LN model is filter-independent, reflecting the global origin of the phenomenon.  The stimulus threshold scales with $\sigma$ and is ${s_{th}}/{\sigma}\approx-\sqrt{2}{\langle v\rangle}/{\sigma}=\sqrt{{2}/{\pi}}$, consistent with the effective operating point being located at the mean subthreshold voltage; the less predictive filter necessarily has a lower threshold (panel C). \textbf{B,D}. Example verification of the conditional rate ansatz in \eqref{RateAnsatz}, shown for $\sigma=6$. (B) STA filter; (D) stochastic linearization filter.   Colors show simulation results for $\big\langle R(t')\big|s_x(t)\big\rangle$; dashed line shows the filter.  At sufficiently large values of $s_x$, the shifted and normalized $\big\langle R(t')\big|s_x(t)\big\rangle$ is approximately the filter (red $\frac{s}{\sigma}=3.3$; orange $\frac{s}{\sigma}=2.2$).  The ansatz breaks down for smaller inputs (green  $\frac{s}{\sigma}=0.66$).}
\label{NonLinHighSigFig}
\end{figure}

We use a couple of tricks to directly derive an asymptotic expression for $R_{\sigma}\!\left[s_x(t)\right]$ for sufficiently large values of $s_x$.  When $s_x$ is sufficiently large, many spikes will occur during the integration window of the filter.  Accordingly, the conditional mean voltage given large $s_x$ must be $\frac{v_{th}-v_r}{2}$ since the voltage trace is either at threshold, reset, or rapidly transitioning between the two.  The mean in Eq. \eqref{MeanVS} becomes
\begin{equation}
\frac{v_{th}-v_r}{2}=\displaystyle v_o+\int_0^{t} \frac{dt'}{\tau}e^{\frac{t'-t}{\tau}}\big(s_x(t)h_x(t-t')-\left(v_{th}-v_r\right)\tau\big\langle R(t')\big|s_x(t)\big\rangle. \label{MeanVSHighSig}
\end{equation}
In this limit, for $\big\langle R(t')\big|s_x(t)\big\rangle$, we take the ansatz
\begin{equation}
\big\langle R(t')\big|s_x(t)\big\rangle\approx\bar{R}_{\sigma}\left(1-\frac{h_x(t-t'+dt)}{\text{max}\!\left(h_x\right)}\right)+R_{\sigma}\!\left[s_x(t)\right]\frac{h_x(t-t'+dt)}{\text{max}\!\left(h_x\right)}. \label{RateAnsatz}
\end{equation}
This ansatz is based on three ideas. First, consistency between the LN and dynamical models requires that the mean conditional rate at $t=t'$ must be the LN model rate, $R_{\sigma}\!\left[s_x(t)\right]$.  Second, for $t'\ll t$, the mean conditional rate must go to $\bar{R}_{\sigma}$, independent of $s_x(t)$.  Third, the time dependence of the mean conditional rate is the same as that of the filter.  This must be approximately true because, at large $s_x(t)$, the dynamics are saturated and the conditional mean voltage must be equal to $\frac{v_{th}-v_r}{2}$ for any large $s_x$ and for times significantly prior to $t$ during the support of the filter---the conditional mean voltage is independent of $s_x$ and is approximately independent of $t$.  We have been unable to derive this from more rigorous grounds but verify its validity numerically in Fig. \ref{NonLinHighSigFig}B. Eqs. \eqref{MeanVSHighSig} and \eqref{RateAnsatz} will cease to hold for values of the filtered stimulus that are small enough so that the conditional mean voltage need not hold steady at $\frac{v_{th}-v_r}{2}$.

The rate estimation function in the limit that $\sigma$ is large compared to $v_{th}-v_o$ and $v_{th}-v_r$, follows from Eqs. \eqref{MeanVSHighSig} and \eqref{RateAnsatz}:
\begin{equation}
\frac{R_{\sigma}\!\left[s_x(t)\right]}{\bar{R}_{\sigma}}=\frac{s_x(t)}{\sigma}\text{max}\!\left(h_x\right)\sqrt{\pi}+\left(1-\frac{\text{max}\!\left(h_x\right)}{h_m\ast h_x}\right)-\left(\frac{v_{th}-2v_o-v_r}{2\sigma}\right)\frac{\text{max}\!\left(h_x\right)\sqrt{\pi}}{h_m\ast h_x}.
\label{NonLinHighSig}
\end{equation}
This formula describes the estimated rate over most of the dynamic range of the neuron. Note that at large $\sigma$, the last term goes to zero and the formula acquires the contrast-invariant form reported in Eq. \eqref{LimitingNonLin}.   Comparisons with numerical results are shown in Fig. \ref{NonLinHighSigFig}.

Similarly, the optimally predictive filter becomes contrast invariant at high $\sigma$.  Using the results in Eqs. \eqref{MeanRateHighSig} and \eqref{PVHighSig} in Eq. \eqref{KslLIF}, the exponential time scale in the continuous time limit is
\begin{equation}
k_{\infty}=1+\frac{\langle v\rangle^2}{\text{Var}\!\left[v\right]}=1+\frac{2}{\pi-2}\approx 2.75. \label{KInfty}
\end{equation}
For $dt=\tau/40$, we find in simulation that $k_{\infty}\approx 2.5$ because the mean firing rate is reduced for finite input correlation time\cite{Brunel2003}. The result for the STA in this limit is given in Eq. \eqref{LimitingSTA}.

\subsubsection*{Higher order moment constraints for contrast gain control}
The shifted moment scaling relations in Eq. \eqref{MomentScaling} are satisfied by the LIF model in the large $\sigma$ limit. We show the first moment explicitly in Eq. \eqref{RateConstraint2}. Here, we argue that higher order moments scale correctly as well, provided the first moment scales correctly. We discuss the LIF model ($f(v)=0$) first.
Each higher order shifted moment involves the averaged sums of products of the different terms in Eq. \eqref{LIFIntegral}.  Heuristically,
\begin{equation}
\left\langle \left(v_{th,\sigma}-v\right)^n\right\rangle =\sigma^n\mu_n \sim \displaystyle \sum_{1\le p< n}  h_m^{(n)}\ast \phi_{i^n} + h_m^{(n)}\ast \phi_{R^n}+h_m^{(n)}\ast \phi_{R^pi^{n-p}} +\sigma^n \mu_{n-p}\mu_p, \notag
\end{equation}
where $h_m^{(n)}\ast \phi$ is $n$-fold convolution with $n$ copies of the membrane filter, $\phi_{i^n}$ is the $n^{th}$-order input autocorrelation,  $\phi_{R^n}$ is the $n^{th}$-order rate autocorrelation, $\phi_{R^pi^{n-p}}$ are the rate-input cross-correlations of combined order $n$, and $\sigma^n \mu_{n-p}\mu_p$ are the products of the lower-order moments.

To understand the scaling of each shifted moment, we identify the scaling of each term. First, for moment $n$, if each lower moment scales correctly, then the contributions from lower moments, $\sigma^n \mu_{n-p}\mu_p$, trivially scale correctly.  Second, at every order, contributions to the shifted moments due to input autocorrelation functions automatically satisfy the scaling relations since the input distribution is Gaussian with standard deviation proportional to $\sigma$.  

Understanding the scaling of the terms involving the rate requires more finesse.  Recall that the mean rate in the contrast invariant regime is linear in $\sigma$.  Looking at the order of each term, contributions from the lowest order rate-input cross-correlations scale correctly:
\begin{equation}
h_m^{(n)}\ast \phi_{Ri^{n-1}}\sim \mathcal{O}(\sigma)\mathcal{O}(\sigma^{n-1})\sim\mathcal{O}(\sigma^n). \notag
\end{equation}
Finally, all contributions from rate autocorrelation functions and rate-input cross-correlations of higher order than one in the rate are always sub-leading relative to the other terms and can be ignored.  We start with the second-order rate autocorrelation function.  Since the integrate-and-fire models are renewal processes, the second order autocorrelation is of the form\cite{GerstnerKistler}
\begin{equation}
\phi_{RR}(t-t')=\bar{R}_{\sigma}\,g(t-t'), \notag
\end{equation}
with integrable $g$. Convolved twice with the membrane filter, the contribution of this moment is $\mathcal{O}(\bar{R})\sim \mathcal{O}(\sigma)$, which is lower order than the $\mathcal{O}(\bar{R}^2)\sim \mathcal{O}(\sigma^2)$ expected from naive power counting.
Higher order powers of $R$ are sub-leading relative to naive power counting as well.  Thus, 
contributions from higher order rate correlation functions to the shifted moments are sub-leading for large $\sigma$, contributing at most:
\begin{equation}
h_m^{(n)}\ast \phi_{R^pi^{n-p}}\sim \mathcal{O}(\bar{R}^{p-1})\mathcal{O}(\sigma^{n-p}) \sim \mathcal{O}(\sigma^{n-1}) \ll  \mathcal{O}(\sigma^n), \notag
\end{equation}
for $n\ge 2$ and $p\ge 2$.  To summarize, when the first shifted moment obeys the perfect gain control constraint, each higher order moment is guaranteed to obey its constraint as well. 

\subsubsection*{Moment constraints for contrast gain control in the  the EIF model}
Results for the EIF model follow similarly: once we have control over the first shifted moment, the others follow for the reasons above. To calculate the first shifted moment, we need the mean subthreshold voltage.  From Eq. \eqref{LIFIntegral}, this is
\begin{equation}
\left\langle v(t)\big|v(t)\le v_{th,\sigma}\right\rangle =v_o  + \displaystyle \int_0^{t}\frac{dt'}{\tau}e^{\frac{t'-t}{\tau}}\left[-\left(v_{s}-v_{r}\right)\tau\left\langle R(t')\big|v(t)\le v_{th,\sigma}\right\rangle+ \left\langle f\big(v(t') \big)\big|v(t)\le v_{th,\sigma}\right\rangle \right].  \notag
 \end{equation}
We can simplify this using the trick introduced preceding Eq. \eqref{ErrorTermEIF}: because of the low pass filtering of the membrane, the contributions of the reset and $f(v)$ when the voltage is above the unstable fixed point are fast and opposite and thus approximately cancel, leaving only the persistent part contributions below $v_{th}$.  This approximation gives
\begin{equation}
\left\langle v(t)\big|v(t)\le v_{th,\sigma}\right\rangle \approx v_o  -\left(v_{th}-v_{r}\right)\tau\bar{R}_{\sigma} +  \displaystyle \int_0^{t}\frac{dt'}{\tau}e^{\frac{t'-t}{\tau}} \left\langle f\big(v(t') \big)\big|v(t)\le v_{th}\right\rangle.  
\label{MeanSubVoltage} \notag
 \end{equation}
The first shifted moment from Eq. \eqref{MomentScaling} is
\begin{equation}
\sigma\mu_1 \approx v_{th,\sigma} - v_o  +\left(v_{th}-v_{r}\right)\tau \bar{R}_{\sigma}- \displaystyle \int_0^{t}\frac{dt'}{\tau}e^{\frac{t'-t}{\tau}}\left\langle f\big(v(t') \big)\big|v(t)\le v_{th}\right\rangle.\notag
\end{equation}
Because $f(v)$ is nonlinear in the voltage, the integrated average will in principle involve terms of all orders in $\sigma$ and so this scaling relation can never be exactly satisfied, even asymptotically, unless $f(v)=0$. However, there are two limits in which we can handle $f(v)$ and both lead to a similar functional result.  

First, for $\Delta\ll v_{th}-v_o$, the contribution from $f(v)$ below threshold because $f(v)$ is small compared to the explicit leak below threshold, and so we can ignore it. 
Also, $v_{th,\sigma}$ grows approximately linearly from $v_{th}$ with $\sigma$, as shown in Eq. \eqref{StochasticThreshold3}. Accounting for the scaling of the stochastic threshold takes $\mu_1$ to a new constant, $\tilde{\mu}_1$, and $v_{th,\sigma}$ to $v_{th}$ in the previous equation. Thus, for the EIF model to show approximately perfect contrast gain control for small $\Delta$, it must also obey a linear rate constraint of the form:
\begin{equation}
\bar{R}_{\sigma}\tau=\sigma \frac{\tilde{\mu}_1}{v_{th}-v_r}-\frac{v_{th}-v_o}{v_{th}-v_r}, \label{RateConstraintEIF0}
\end{equation}
where $\tilde{\mu}_1$ is determined by the model parameters; for the LIF model ($\Delta\rightarrow 0$), comparison with Eq. \eqref{MeanRateHighSig} gives $\tilde{\mu}_1=1/\sqrt{\pi}$. The stochastic threshold drops out of this expression because contrast gain control is a global phenomenon and does not depend on the exact definition of spike times. 

For $v_{th}-v_o\lesssim\Delta\ll v_{s}-v_o$, $f(v)$ is no longer small.\footnote{The limit of $\Delta\sim v_s-v_o$ is unphysiological because the excitability of $f(v)$ and separation between integration and spiking disappears: the model becomes equivalent to a leakless integrator with a reflecting boundary at $v_o$ and absorbing threshold at $v_s$.}   For hyperpolarized voltages, $v\lesssim v_o-\Delta$, $f(v)$ below threshold is primarily linear and increases the leak, as shown in Eq. \eqref{TauLeak}.  The first shifted moment can be re-expressed in terms of the hyperpolarized leak time constant, $\tau_L$:
\begin{equation}
\sigma\mu_1 \approx v_{th,\sigma} - v_o  +\left(v_{th}-v_{r}\right)\tau_L \bar{R}_{\sigma}- \displaystyle \int_0^{t}\frac{dt'}{\tau_L}\frac{e^{\frac{t'-t}{\tau_L}}}{1-f'_{-\infty}}\left\langle f\big(v(t') \big)-(v-v_o)f'_{-\infty}\big|v(t)\le v_{th}\right\rangle,\notag
\end{equation}
where $f'_{-\infty}$ is the limiting slope of $f(v)$ for large negative voltages. For larger $\Delta$, $\tau_L\rightarrow 0$, so the time integral is supported only near $t'=t$.  Also, $f(v)\rightarrow v_{th}-v_o$ for  $v_o-\Delta \lesssim v \lesssim v_{th}+\Delta$.  Thus, in the limit, for $\sigma \sim v_{th}-v_o$, we have:
\begin{equation}
\displaystyle \int_0^{t}\frac{dt'}{\tau_L}\frac{e^{\frac{t'-t}{\tau_L}}}{1-f'_{-\infty}}\left\langle f\big(v(t') \big)-(v-v_o)f'_{-\infty}\big|v(t)\le v_{th}\right\rangle\approx v_{th}-v_o,\notag
\end{equation}
and the first moment constraint becomes
\begin{equation}
\sigma\mu_1 \approx v_{th,\sigma} - v_{th}  +\left(v_{th}-v_{r}\right)\tau_L \bar{R}_{\sigma}.\notag
\end{equation}
After accounting for the linear scaling of the stochastic threshold, the mean rate constraint derived from the voltage distribution scaling for larger $\Delta$ is
\begin{equation}
\bar{R}_{\sigma}\tau_L=\sigma \frac{\tilde{\mu}_1}{v_{th}-v_r}. \label{RateConstraintEIFInf}
\end{equation}
Comparison of Eqs. \eqref{MeanRateConstraint} and \eqref{RateConstraintEIFInf} shows that perfect contrast gain control is guaranteed for $\Delta\sim 1$. We discuss this in more detail in Section: {\em Contrast gain control in the EIF model}.

The derived constraints on the rate based on the scaling of the voltage distribution can be summarized across $\Delta$ as a linear constraint with variable intercept:
\begin{equation}
\bar{R}_{\sigma}\tau=\sigma \frac{\tilde{\mu}_1}{v_{th}-v_r}-\nu_1\left(\frac{v_{th}-v_o}{v_{th}-v_r}\right), \label{RateConstraintEIF}
\end{equation}
where $\nu_1$ is a constant between zero and one and is determined by the model parameters.

% Do NOT remove this, even if you are not including acknowledgments
\section*{Acknowledgments}
We thank Julijana Gjorgjieva for her insights into contrast invariance in the LIF model and her careful prodding and questioning regarding this work.  We thank Sean Trettel for helpful analysis regarding the maximally informative filter.

\section*{Funding information}
This work has been supported by a McKnight Scholar Award in the Neurosciences, and by NSF grant $\# 0928251$ (Advancing Theory in Biology).

\section*{Author Contributions}
Project conception and writing: MF and AF.  Mathematical framework, development, and simulations: MF.

% can submit .bib and don't have to embed refs
%\section*{References}
%\bibliography{/Users/Famulare/Dropbox/Writing/Famulare_refs}

\end{document}